\documentclass[aps]{revtex4-1}
\usepackage{graphicx}
\usepackage{amsmath}
\usepackage{amssymb}
\usepackage{tikz}
\usepackage{float}
\usepackage{sidecap}
\usepackage[caption = false]{subfig}
\begin{document}
\title{\bf Mass Spectra of Heavy-Light Mesons in Heavy Hadron 
Chiral Perturbation Theory}
\author{Mohammad H. Alhakami}
\affiliation{Nuclear Science Research Institute\\
KACST,  P.O. Box 6086, Riyadh 11442, Saudi Arabia}
\date{\today}
\begin{abstract}
We study the masses of the low-lying charm and bottom mesons within the framework of heavy hadron chiral perturbation theory ($\mathrm{HHChPT}$). We work to third order in the chiral expansion,
 where meson loops contribute. In contrast to previous approaches, we use physical meson masses in evaluating these loops. This ensures that their imaginary parts are
 consistent with the observed widths of the $D$-mesons. The lowest odd- and even-parity, strange and nonstrange charm mesons provide enough constraints to determine only
certain linear combinations of the low-energy constants in the effective Lagrangian. We comment on how lattice QCD could provide further information to disentangle
 these constants.
Then we use the results from the charm sector to predict the spectrum of odd- and even-parity of the bottom mesons.
The predicted masses from our theory are
in good agreement with experimentally measured masses
for the case of the odd-parity sector.
For the even-parity sector, the $B$-meson states have not yet been observed; thus,
our results provide useful
information for  experimentalists  investigating such states. 
The near degeneracy of nonstrange and strange scalar $B$ mesons
is confirmed in our predictions using $\mathrm{HHChPT}$.
We show why previous approaches
of  using
$\mathrm{HHChPT}$ in studying the mass degeneracy in the scalar states
of charm and bottom meson sectors
gave unsatisfactory results.
\end{abstract}


\maketitle

\section {Introduction}
The masses and widths of the low-lying charm mesons are now rather well 
determined experimentally, in the odd- and even-parity, strange and nonstrange 
sectors (for summaries, see Refs.~\cite{pdg12,cdgn12}). The patterns of the masses 
and interactions of these mesons are governed by two approximate symmetries: the 
spin symmetry of the heavy quark and the $SU(3)_L\times SU(3)_R$ chiral symmetry 
of the light quarks. 
Both symmetries can be incorporated in a single framework using heavy-hadron
chiral perturbation theory ($\mathrm{HHChPT}$), an effective field theory 
for the interactions of a meson containing a single heavy quark \cite{4,cas97,FalkLuke,PCh,wise, Julish}.

Within this theory, the masses of the low-lying odd- and even-parity $D$ mesons 
have been studied, including one-loop chiral corrections \cite{ms05,absu07}.
The chiral Lagrangian at this, third, order contains a number of unknown 
low-energy constants (LECs). These cannot be determined uniquely from 
experimental data on the meson spectrum because their number exceeds the number 
of low-lying mesons. Mehen and Springer \cite{ms05} and Ananthanarayan 
\textit{et al.}~\cite{absu07} fitted expressions that depend nonlinearly on these
constants and found multiple solutions, often with quite different numerical 
values for them. As a result, no clear pattern emerged from these fits.

In this paper, we use a different approach to fit these parameters to
 remove these ambiguities and provide a clearer picture. 
The key difference from previous work \cite{ms05,absu07} is that we use the 
physical values of the charm meson masses in evaluating the chiral loops. 
One important consequence of this is to put thresholds at the correct energies 
relative to the masses of unstable particles and hence to ensure that the 
imaginary parts of the loops are correctly related to the observed decay widths 
of the heavy mesons.
A second consequence is that the parameters -- the LECs -- appear only in 
the tree-level contributions to the masses. This allows us to determine uniquely
eight linear combinations of the LECs from the experimental masses. These eight
parameters cannot be further disentangled into the individual LECs using the
experimental spectrum alone.
By using the experimental masses in the loops, we generate terms that are of order
higher than third order in the chiral expansion. These include divergences that
we cannot cancel using counterterms in our Lagrangian. We use the $\beta$-functions 
associated with these uncontrolled higher-order contributions to provide an estimate
of the theoretical errors introduced by our approach.
Another, more technical, difference from previous work on $\mathrm{HHChPT}$ is that 
 we have used corrected expressions for the chiral loop functions, in contrast to  the expressions presented in \cite{ms05,cas97} which
use an inconsistent renormalization scheme.

The results from the charm meson sector are used to predict the masses of the full set
of the low-lying $B$-meson states. The predicted masses from our theory of the ground states are in good
agreement with the well-determined masses. The first set of excited $B$ meson states has not yet been observed; thus our results can be used to provide useful information for experimentalists investigating such states.
The near degeneracy of scalar $B$ meson states-- the mass of the nonstrange scalar $B$-meson is similar to that of strange one-- is confirmed in our predictions using $\mathrm{HHChPT}$.
Our results are at variance with those in Ref. \cite{bmeson}.
 We will show why the previous 
 studies of the near mass degeneracy in the scalar $D$- and $B$-meson sectors using the approach of $\mathrm{HHChPT}$ led to unsatisfactory results.
 
This paper is organized as follows. In Sec. II, the heavy-hadron chiral Lagrangian we 
use is briefly reviewed. In Sec. III, we present the resulting expression for the meson 
masses. Since the number of the LECs exceeds the number of obervables, the LECs are grouped into eight linear combinations
that are equivalent to the number of observables. 
In Sec. IV, we use the $D$-meson spectrum to fit these parameters.
The results from the charm meson spectrum
are then used in Sec. V to predict the masses of the low-lying bottom meson states. The summary is given
in Sec. VI.

\section{Heavy-hadron chiral Lagrangian}
Our starting point is the same effective Lagrangian that was used in 
Refs.~\cite{ms05,absu07}. We give a brief outline of it here; more details can 
be found in those papers and the review by Casalbuoni \textit{et al.}~\cite{cas97}.
In the heavy quark limit, systems with a single heavy 
quark respect heavy-quark spin symmetry, forming degenerate multiplets independent 
of the spin orientation of the quark. The lowest multiplet of charm mesons 
consists of the pseudoscalar ground states, $D^0$, $D^+$ and $D^+_s$, and their 
vector first excited states, $D^{*0}$, $D^{*+}$ and $D^{*+}_s$. These can be
conveniently described by the effective field,
\begin{equation}
\mathcal{H}_a=  \frac{1+v\llap/}{2}\left(H_a^\mu\gamma_\mu-H_a\gamma_5\right),
\end{equation}
where the fields $H_a$ annihilate the pseudoscalar particles and $H_a^\mu$ annihilate
the vector ones. Here, the flavor index $ a= 1,2,3$ denotes states with up, down, 
and strange quarks, respectively. 
The first excited multiplet has the opposite parity and consists of scalar, 
$D^0_0$, $D^+_0$ and $D^+_{0s}$, and axial-vector mesons, $D^{0'}_1$, $D^{1'}_1$
and $D^{0'}_{1s}$. These can be
  described by the effective field, 
\begin{equation}
 \mathcal{S}_a= \frac{1+v\llap/}{2}\left(S^{\mu}_a\gamma_{\mu}\gamma_5-S_a\right),
\end{equation}
where the fields $S_a$ and $S_a^\mu$  annihilate the scalar and axial-vector particles, respectively. 

The other ingredient of the theory is the approximate $SU(3)_L\times SU(3)_R$
chiral symmetry of QCD. This is embodied by fields describing the lightest 
strongly interacting particles, $\pi$, $K$ and $\eta$, which are approximately 
the Goldstone bosons of this hidden symmetry. These can be represented by  
 the matrix field $U(x)=\exp({\rm i}\sqrt{2}\phi(x)/f)$ where  $\phi(x)$ is given by
\begin{equation}
\phi(x)=\left (\begin{array}{ccc}\frac{1}{\sqrt{2}}\pi^0+\frac{1}{\sqrt{6}}\eta&\pi^+&K^+ \\ \pi^- & -\frac{1}{\sqrt{2}}\pi^0+\frac{1}{\sqrt{6}}\eta&K^0\\
K^-&\bar{K}^0&-\sqrt{\frac{2}{3}}\eta \end{array}\right).
\end{equation}
In our conventions,
we use the physical value of the pion decay constant $f=92.4\,\mathrm{MeV}$.
It is different from the ones used by Wise in \cite{wise}
in which  $f=135\,\mathrm{MeV}$ was used.  
Thus,
one has to replace $f$ in \cite{wise} by $\sqrt{2}f$ to account for
different conventions.
The lowest-order Lagrangian for the light mesons is
\begin{equation}
\mathcal{L}_m=\frac{f^2}{4}\mathrm{Tr}\left(\partial_\mu U\partial^\mu U^\dagger\right)+\frac{f^2\,B_0}{2}\mathrm{Tr}\left(m_q\, U^{\dag }+U\,m_q^{\dag}\right),
\end{equation}
where 
the coefficient $B_0$ is related to the pion decay constant and the quark condensate 
of light quark flavors \cite{sch03}. The light quark mass matrix is given by $m_q=\mathrm{diag}(m_u,m_d,m_s)$. 

We take as our low-energy scales, generically denoted by $Q$, the masses and momenta 
of the Goldstone bosons and the splittings between the four lowest states of the 
$D$ mesons introduced above. The relevant expression of the heavy-hadron chiral Lagrangian up to order $Q^3$ is \cite{4,ms05}
\begin{equation}
\begin{split}
{\mathcal{L}}_H=&-\mathrm{Tr}[\overline{\mathcal{H}_a}\left(i v\cdot D_{ba} 
-\delta_H \delta_{ab}\right) \mathcal{H}_b]
+\mathrm{Tr}[\overline{\mathcal{S}}_a\left(i v\cdot D_{ba} 
-\delta_S \delta_{ab}\right) \mathcal{S}_b]\\
&+ g \mathrm{Tr}[\overline{\mathcal{H}}_a\mathcal{H}_b {u\llap/}_{ba} \gamma_5]
+g^{\prime}  \mathrm{Tr}[\overline{\mathcal{S}}_a\mathcal{S}_b {u\llap/}_{ba}\gamma_5]
+ h \mathrm{Tr}[\overline{\mathcal{H}}_a\mathcal{S}_b {u\llap/}_{ba}\gamma_5 + \mbox{h.c.}]\\
&-\frac{\Delta_H}{8}\mathrm{Tr}[\overline{\mathcal{H}}_a\sigma^{\mu \nu}\mathcal{H}_a\sigma_{\mu\nu}]
+\frac{\Delta_S}{8}\mathrm{Tr}[\overline{\mathcal{S}}_a\sigma^{\mu \nu}\mathcal{S}_a\sigma_{\mu\nu}]\\
&+a_H \mathrm{Tr}[\overline{\mathcal{H}}_a\mathcal{H}_b] m^{\xi}_{ba}-a_S \mathrm{Tr}[\overline{\mathcal{S}}_a\mathcal{S}_b] m^{\xi}_{ba} + \sigma_H \mathrm{Tr}[\overline{\mathcal{H}}_a\mathcal{H}_a] m^{\xi}_{bb}-\sigma_S \mathrm{Tr}[\overline{\mathcal{S}}_a\mathcal{S}_a] m^{\xi}_{bb}\\
&-\frac{\Delta^{(a)}_H}{8}\mathrm{Tr}[\overline{\mathcal{H}}_a\sigma^{\mu\nu}\mathcal{H}_b \sigma_{\mu\nu}]m^{\xi}_{ba}
+\frac{\Delta^{(a)}_S}{8}\mathrm{Tr}[\overline{\mathcal{S}}_a\sigma^{\mu\nu}\mathcal{S}_b\sigma_{\mu\nu}]m^{\xi}_{ba}\\
&-\frac{\Delta^{(\sigma)}_H}{8}\mathrm{Tr}[\overline{\mathcal{H}}_a\sigma^{\mu\nu}\mathcal{H}_a\sigma_{\mu\nu}]m^{\xi}_{bb}
+\frac{\Delta^{(\sigma)}_S}{8}\mathrm{Tr}[\overline{\mathcal{S}}_a\sigma^{\mu\nu}\mathcal{S}_a\sigma_{\mu\nu}]m^{\xi}_{bb},\label{M3}
\end{split}
\end{equation}
where  the covariant derivative is defined as $D^\mu_{ba}=\partial^{\mu}_{ba}+\frac{1}{2}(\xi^\dag \, \partial^\mu \xi+\xi\, \partial^\mu \xi^\dag)_{ba}$,
$\xi(x)=\sqrt{U(x)}$. The factors $\delta_H$ and $\delta_S$ are the residual masses of  the effective fields $\mathcal{H}_a$ and $\mathcal{S}_a$, respectively.
The coupling constant $g$ ($g^{\prime}$) measures the strength of transitions within
odd- (even-)parity charm meson states. 
The strength of transitions between odd- and even-parity states
is measured by the coupling constant $h$.
The axial vector field is $u^{\mu}_{ba}=\frac{i}{2}( \xi^\dag \, \partial^\mu \xi-\xi\, \partial^\mu \xi^\dag)_{ba}$.
The hyperfine splittings of the $D$-meson states are measured by ($\Delta$, $\Delta^{(a)}$, $\Delta^{(\sigma)}$).
These coefficients manifestly vanish in the heavy quark limit. 
The quark mass matrix which breaks chiral symmetry is defined as $m^{\xi}_{ba}=\frac{1}{2}( \xi \, m_q \xi+ \xi^\dag m_q \xi^\dag)_{ba}$.
The coefficients ($a$, $\sigma$) present in the chirally breaking terms are dimensionless.

According to our power counting, the terms in the first three lines in Eq.~\eqref{M3}
are all of order $Q^1$. These include terms, in the third line, that break the 
heavy-quark spin symmetry. Since, at leading order, the quark masses are 
proportional to the squares of the masses of the Goldstone bosons, the 
chiral-symmetry breaking terms in the fourth line are of order $Q^2$. 
The final terms which break both chiral and heavy-quark spin symmetries are of 
order $Q^3$.
These terms  are required to cancel the infinite parts resulting from regularization and renormalization of the loop diagrams, note that all diagrams are of order $Q^3$.
\begin{figure}[h!]
\begin{center}
\begin{tikzpicture}[domain=-2:10,scale=0.9]
\draw[-,thick,line width=1pt] (6,2)--(7,2)   (6.2,2) node[above]{$H$};
\draw[double,line width=1pt] (7,2)--(9,2) (8,2) node[above]{$S$};
\draw [dashed,line width=1pt] (7,2) arc (180:0: 1cm) (8,3) node[above]{$\pi$,\,$K$,\,$\eta$};
\draw[-,thick,line width=1pt] (9,2)--(10,2)  (9.8,2) node[above]{$H$};
\draw[-,thick,line width=1pt] (1,2)--(2,2)   (1.2,2) node[above]{$H$};
\draw[-,thick,line width=1pt] (2,2)--(4,2) (3,2) node[above]{$H$};
\draw[-,thick,line width=1pt] (4,2)--(5,2)  (4.8,2) node[above]{$H$};
\draw [dashed,line width=1pt] (2,2) arc (180:0: 1cm) (3,3) node[above]{$\pi$,\,$K$,\,$\eta$};
\draw[double,line width=1pt] (6,0)--(7,0)   (6.2,0) node[above]{$S$};
\draw[double,line width=1pt] (7,0)--(9,0) (8,0) node[above]{$S$};
\draw [dashed,line width=1pt] (7,0) arc (180:0: 1cm) (8,1) node[above]{$\pi$,\,$K$,\,$\eta$};
\draw[double,line width=1pt] (9,0)--(10,0)  (9.8,0) node[above]{$S$};
\draw[double,line width=1pt] (1,0)--(2,0)   (1.2,0) node[above]{$S$};
\draw[-,thick,line width=1pt] (2,0)--(4,0) (3,0) node[above]{$H$};
\draw[double,line width=1pt] (4,0)--(5,0)  (4.8,0) node[above]{$S$};
\draw [dashed,line width=1pt] (2,0) arc (180:0: 1cm) (3,1) node[above]{$\pi$,\,$K$,\,$\eta$};
\end{tikzpicture}
\caption{The self-energy diagrams for the ground-state fields $H$ and the excited-state fields $S$.}
\label{figgraph}
\end{center}
\end{figure}
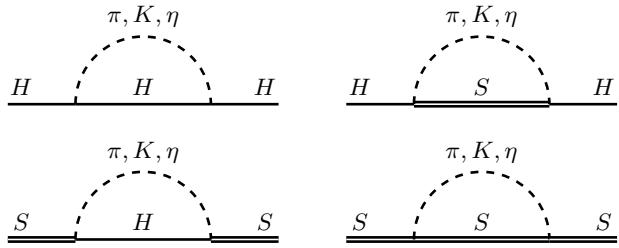

\section{Mass formula of the charm mesons}
The full contributions to the physical mass can be obtained by adding  the tree-level contributions to the one-loop corrections $\Sigma_{D^{(*)}}$ as  
\begin{equation}
\begin{split}\label{mesonspectra}
m_{H_a}&= \delta_H+a_H m_a+\sigma_H \overline{m}-\frac{3}{4}(\Delta_H+\Delta^{(a)}_H m_a+\Delta^{(\sigma)}_H \overline{m})+\Sigma_{H_a},\\[2ex]
m_{H^*_a}&= \delta_H+a_H m_a+\sigma_H \overline{m}+\frac{1}{4}(\Delta_H+\Delta^{(a)}_H m_a+\Delta^{(\sigma)}_H \overline{m})+\Sigma_{H^*_a},\\[2ex]
m_{S_a}&= \delta_S+a_S m_a+\sigma_S \overline{m}-\frac{3}{4}(\Delta_S+\Delta^{(a)}_S m_a+\Delta^{(\sigma)}_S \overline{m})+\Sigma_{S_a},\\[2ex]
m_{S^*_a}&= \delta_S+a_S m_a+\sigma_S \overline{m}+\frac{1}{4}(\Delta_S+\Delta^{(a)}_S m_a+\Delta^{(\sigma)}_S \overline{m})+\Sigma_{S^*_a},
\end{split}
\end{equation}
where we use the notation of Ref. \cite{ms05}.  Here, we work in the isospin limit ($m_u=m_d=m_1$) where $\overline{m}=2\,m_1+m_3$
and $m_a=(m_1,m_1,m_3)$. 
The Feynman diagrams of the one-loop corrections $\Sigma_{D^{(*)}}$ to the masses of $D$-mesons are shown in Fig.\ref{figgraph}.
The resulting explicit expressions for  the self energies of the charm mesons  are given in Appendix A.
In our work,  the residual masses $m_{D^{(*)}}$ are measured from the nonstrange spin-averaged $H$ mass, $(m_{H_1}+3 m_{H_1^*})/4$.

The existing coefficients  in Eq.~\eqref{mesonspectra} can be either determined from  experiments or from
lattice fit. 
In Refs. \cite{ms05,absu07}, the authors fitted the above expressions which depend nonlinearly on these coefficients and found multiple solutions,
 often with quite different numerical values for them. As a result, no clear pattern emerged from these fits.
This is because the number of these coefficients 
exceeds the number of  experimentally known charm meson masses.
Thus, getting unique numerical values of the coefficients is impossible.
Here, we attempt  to remove this ambiguity by following a different approach to fit these coefficients.
We use the physical values of the masses  in evaluating the chiral loops.
As a consequence, the energy of any unstable particle is placed correctly relative to the decay threshold, and the imaginary part of
the loop integral can be related to the experimental decay width.
The second effect is to reduce the number of unknown coefficients in comparison with the current experimental data on charm meson masses.
Masses at tree level depend only on certain linear  combinations of LECs. By using physical masses in chiral loops, 
the masses still depend linearly on these combinations.
Therefore, one can express these combination of LECs directly in terms of the physical masses and loop integrals.

The  procedure of combining the LECs is performed according to the symmetry patterns of the charm mesons. 
In this manner, the constructed parameters can be uniquely determined by using available experimental values of the meson masses and widths.
The  parameters that respect flavor symmetry are
\begin{equation}
\begin{split}\label{startwik}
&\eta_H=\delta_H+(\frac{a_H}{3}+\sigma_H)\, \overline{m},~~\xi_H=\Delta_H+(\frac{\Delta^{(a)}_H}{3}+\Delta^{(\sigma)}_H)\, \overline{m},\\[2ex]
&\eta_S=\delta_S+(\frac{a_S}{3}+\sigma_S)\, \overline{m},~~\xi_S=\Delta_S+(\frac{\Delta^{(a)}_S}{3}+\Delta^{(\sigma)}_S)\, \overline{m},
\end{split}
\end{equation}
where $\delta_{H;S}$ and  $\Delta_{H;S}$ respect chiral symmetry, but the other terms contain the average of the quark masses $\overline{m}$ which breaks it.
The parameters left after constructing  $\eta_H$, $\eta_S$, $\xi_H$,  and $\xi_S$ are
\begin{equation}
\begin{split}
&L_{H}=(m_3-m_1)\,a_H,~~T_{H}=(m_3-m_1)\,\Delta^{(a)}_H,\\[2ex]
&L_{S}=(m_3-m_1)\,a_S,~~T_{S}=(m_3-m_1)\,\Delta^{(a)}_S.
\end{split}
\end{equation}
The combinations $L_{H;S}$ and $T_{H;S}$ break flavor symmetry, and the latter also breaks  spin symmetry.
In terms of these linear combinations, the masses can be written as 
\begin{equation}
\begin{split}
&m_{H_a}=\eta_H-\frac{3}{4}\xi_H+\frac{\alpha_a}{3} L_{H}+\frac{\beta_a}{2} T_{H}+\Sigma_{H_a},\\[2ex]
&m_{H^*_a}=\eta_H+\frac{1}{4}\xi_H+\frac{\alpha_a}{3} L_{H}+\frac{\beta^*_a}{2} T_{H}+\Sigma_{H^*_a},\\[2ex]
&m_{S_a}=\eta_S-\frac{3}{4}\xi_S+\frac{\alpha_a}{3} L_{S}+\frac{\beta_a}{2} T_{S}+\Sigma_{S_a},\\[2ex]
&m_{S^*_a}=\eta_S+\frac{1}{4}\xi_S+\frac{\alpha_a}{3} L_{S}+\frac{\beta^*_a}{2} T_{S}+\Sigma_{S^*_a},
\end{split}
\end{equation}
where $\alpha_a$ and $\beta^{(*)}_a$ are $\alpha_1=-1$, $\alpha_3=2$, $\beta_1=1/2$, $\beta_3=-1$, 
$\beta^*_1=-1/6$, and $\beta^*_3=1/3$.  
Now, the number of parameters,  $\xi_{H;S}$, $\eta_{H;S}$, $L_{H;S}$, and $T_{H;S}$ is $8$, which is
 equal to the number of observed low-lying $D$-meson states.

\section{Determination of low-energy constants}
The numerical values of the parameters  ($\xi_{H;S}$, $\eta_{H;S}$, $L_{H;S}$, $T_{H;S}$) will be given in this part.
In our fitting, the physical masses and the coupling constants extracted from the well-measured widths are used.
The used meson masses are two masses of the ground-state nonstrange mesons in the isospin limit and  four masses of strange mesons from both sectors,
see Table \ref{table:1.4bb}.
The  excited  nonstrange mesons are reported with the large uncertainties. In this case, we did not take the isospin average and instead the
masses of the neutral heavy mesons ($m_{D^0_0}=2318\pm29~\mathrm{MeV}$ \cite{pdg12}, \, $m_{D^{0'}_1}=2427\pm36~\mathrm{MeV}$ \cite{19}) 
are chosen due to their relatively small errors in comparison with the excited charged mesons \cite{be03,19,cl0,cl1,fo04,ba03,pdg12}.
The masses of the Goldstone particles used here are ($m_{\pi}=140~\mathrm{MeV}$, $m_{K}=495~\mathrm{MeV}$, and $m_{\eta}=547~\mathrm{MeV}$).
The calculations are performed at the physical values of pion decay constant $f=92.4~\mathrm{MeV}$ and of the coupling constants $g$ and $h$ that 
are extracted from the strong decay widths $g=0.64\pm0.075$ and $h=0.56\pm0.04$; for details, see \cite{cdgn12}.
The renormalization scale $\mu$ is chosen to be the average of the pion and kaon masses $\mu=317~\mathrm{MeV}$.

\begin{table}[h!]
\def\arraystretch{1.5}
\begin{center}
\begin{tabular}{|c|c|c||c|c|c||c|c|c|}
\hline
Name&$J^{p}$&Mass (MeV)&Name&$J^{p}$&Mass (MeV)&Name&$J^{p}$&Mass (MeV)\\  \hline  \hline 
$D^0$&$0^-$&$1864.84\pm0.05$&$D^\pm$&$0^-$&$1869.61\pm0.09$&$D^\pm_s$&$0^-$&$1968.30\pm0.10$\\  \hline
$D^{*0}$&$1^-$&$2006.97\pm0.08$&$D^{*\pm}$&$1^-$&$2010.27\pm0.05$&$D^{*\pm}_s$&$1^-$&$2112.1\pm 0.4$\\  \hline
$D_0^0$&$0^+$&$2318\pm 29$&$D^{\pm}_0$&$0^+$&$...$&$D_{s0}^{*\pm}$&$0^+$&$2317.7 \pm 0.6$\\  \hline
$D_1^{0\prime}$&$1^+$&$2427 \pm36$&$D^{\pm\prime}_{1}$&$1^+$&$...$&$D^{\pm\prime}_{s1}$&$1^+$&$2459.5\pm 0.6$\\  \hline
\end{tabular}
\caption{The listed charm meson states have been used in our fitting. $J^p$ is the angular momentum and parity of the meson.
In our fitting, the masses of $H_1$ ($H^*_1$) are obtained by taking the isospin average of
$D^0$ and $D^\pm$ ($D^{*0}$ and $D^{*\pm}$); for details please refer to the text. All masses are taken from the Particle Data Group \cite{pdg12}
except the mass of the excited neutral nonstrange meson $D_1^{0\prime}$, which is reported by the BELLE collaboration \cite{19}.}
\label{table:1.4bb}
\end{center}
\end{table}

The chiral-loop functions are fed with the difference of the physical masses of the charm mesons. Thus, the uniquely
determined values of the parameters include contributions from terms beyond the loop order.
Since these higher-order terms have not been considered in the chiral Lagrangian, their $\mu$ dependence cannot 
be canceled by existing coefficients. So, beta functions of the parameters are defined in order
to estimate how much higher-order terms donate to the central values of those parameters.
The resulting numerical values of the parameters which inhabit the odd-parity sector   
 are 
 \begin{equation}
\begin{split}\label{Odd}
 &\eta_H=171.57\pm44\pm5 \, \mathrm{MeV},~~\xi_H=150.95\pm5\pm5 \,\mathrm{MeV},\\[2ex]
&L_{H}=242.71\pm40\pm18\,\mathrm{MeV},~~T_{H}=-52.21\pm18\pm15\,\mathrm{MeV},
\end{split}
\end{equation}
where the first uncertainty is  the experimental error associated with physical masses of charm mesons and the
second uncertainty is the theoretical error that we have estimated from the $\beta$-functions.
\begin{figure}
\subfloat[ ]{\includegraphics[width = 3in]{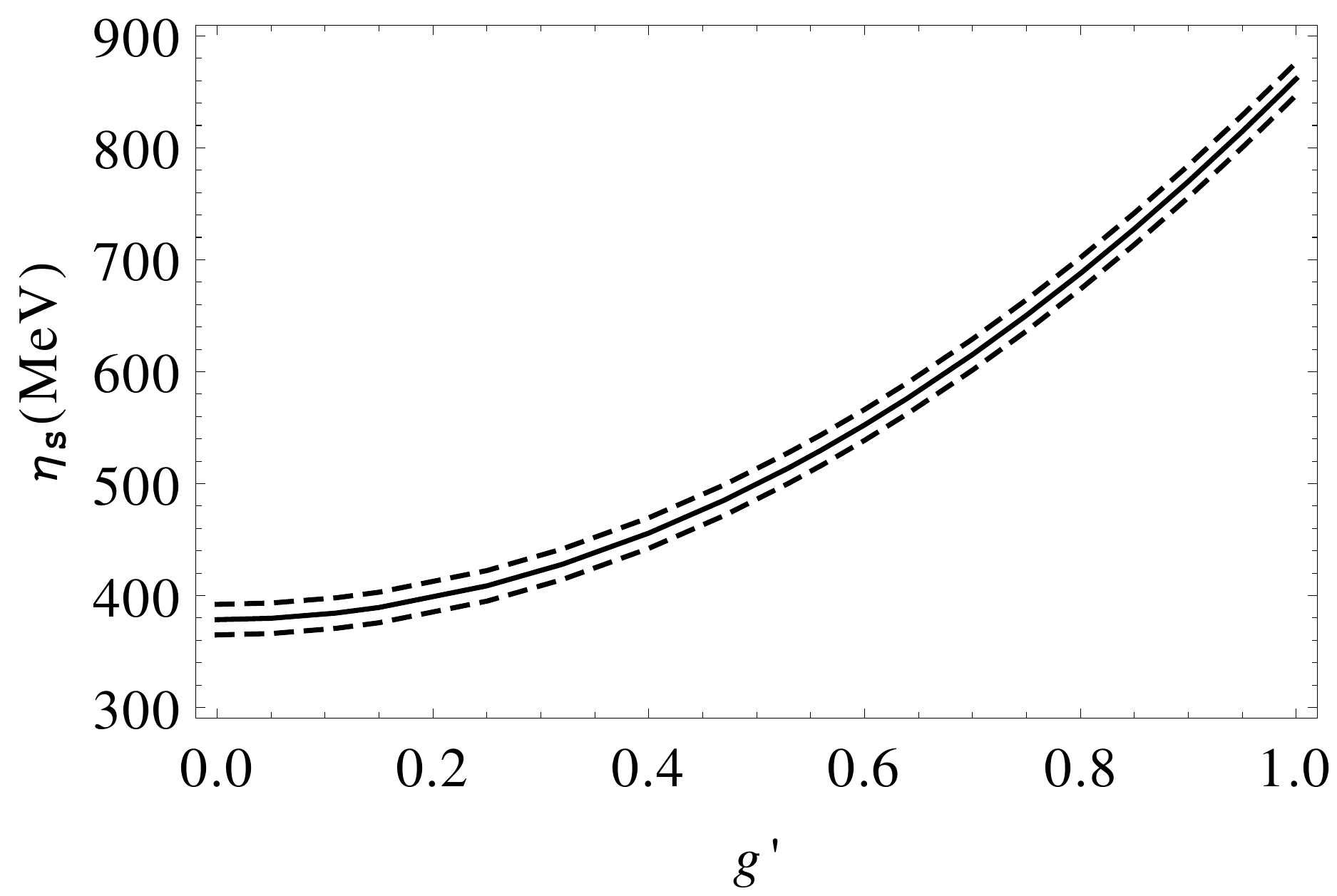}} 
\subfloat[ ]{\includegraphics[width = 3in]{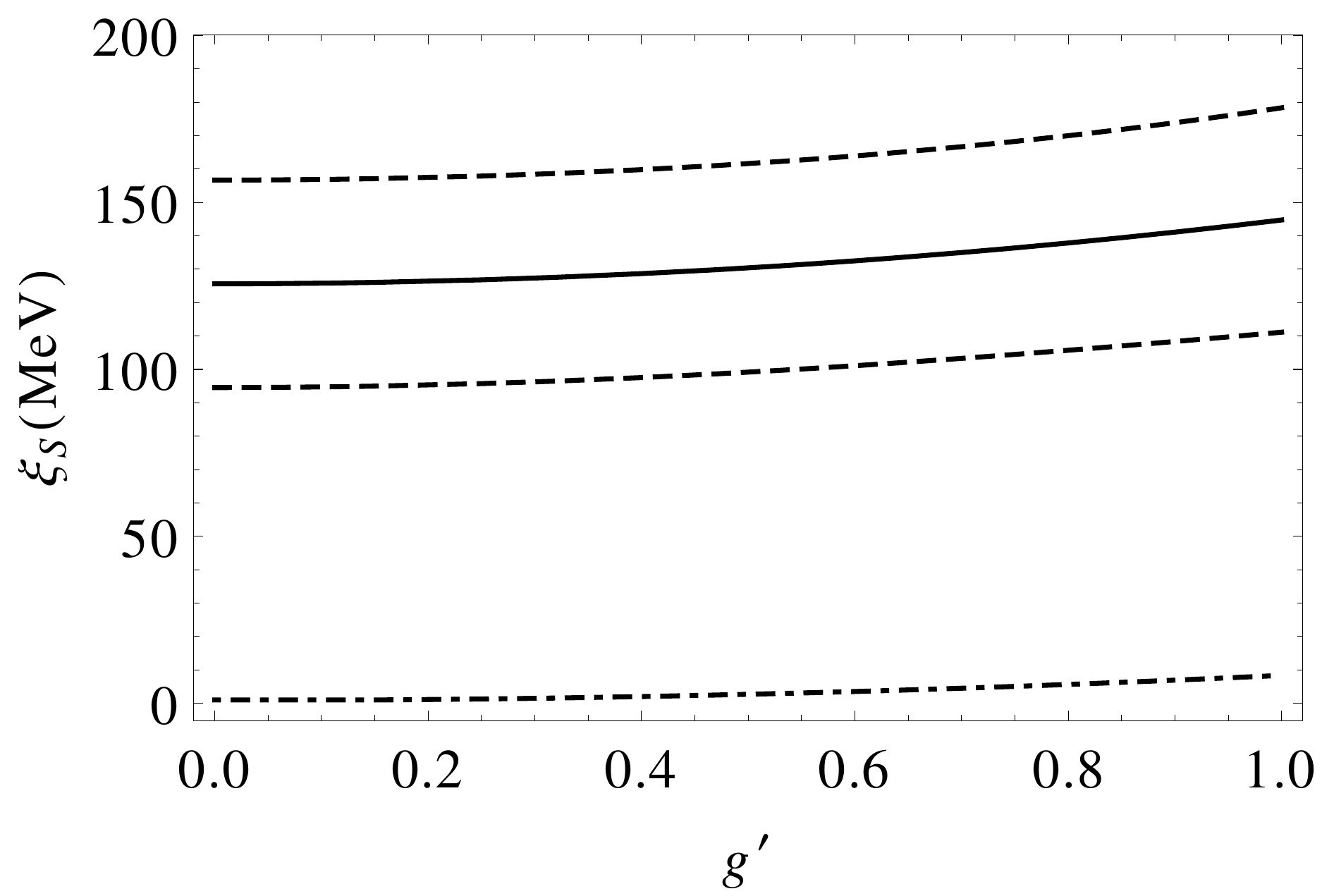}}\\
\subfloat[ ]{\includegraphics[width = 3in]{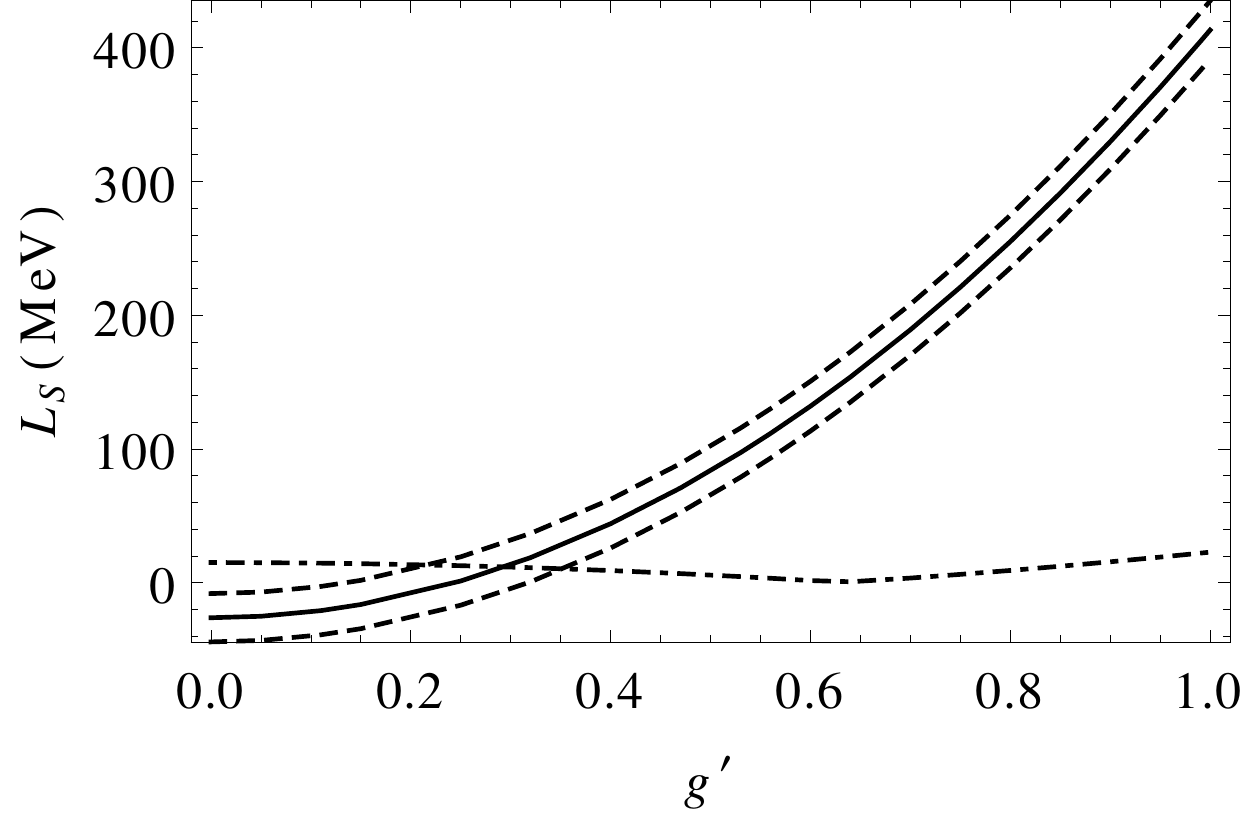}}
\subfloat[ ]{\includegraphics[width = 3in]{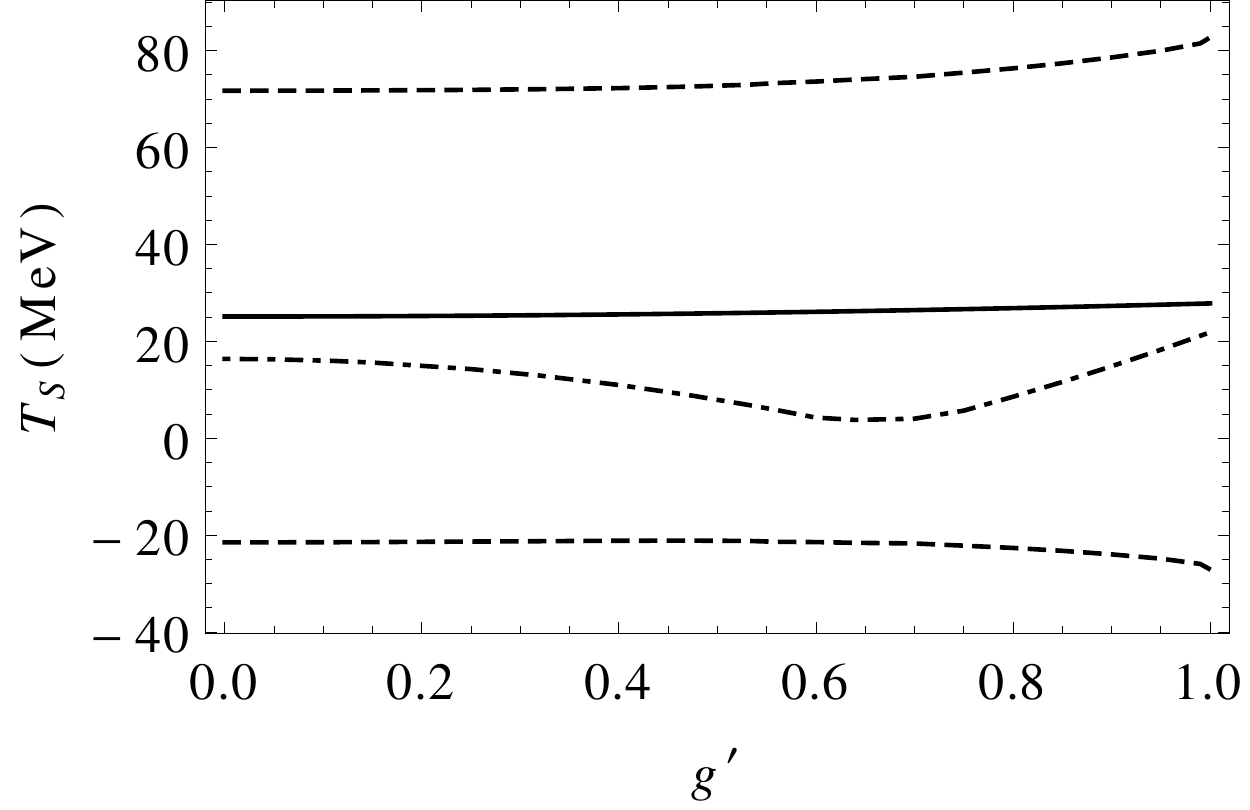}} 
\caption{Variation of 
(a) $\eta_{S}$, (b) $\xi_{S}$, (c) $L_{S}$, and (d) $T_{S}$ with $g^{\prime}$.
The experimental uncertainties are shown by dashed lines surrounding the
central values, and an estimate theoretical
uncertainty is shown by dotted-dashed line.
The theoretical uncertainty of the parameter $\eta_S$ is a constant $\pm 5\, \mathrm{MeV}$.}
\label{fig3}
\end{figure}

The situation for the even-parity parameters is different because the coupling constant $g^{\prime}$ is not determined experimentally.
Since the value of the odd-parity coupling constant $g$ is $0.64$, it is plausible to consider values for  $g^{\prime}$ in the range $0$ to $1$.
The correlations between $g^{\prime}$ and $\eta_S$, $\xi_S$, $L_{S},T_{S}$ are
shown in Fig.~\ref{fig3}. 
The plots also show the associated experimental and theoretical errors.

Experimental information is not sufficient to separate the combinations of the LECs into pieces that respect and break chiral symmetry, which limits their usefulness for applications
to other observables. Lattice  $\mathrm{QCD}$ calculations would be required to perform further separations of terms.
For example,
lattice results on the charm meson spectroscopy undertaken in Refs. \cite{Canada,Japan}
can be used
to disentangle chirally symmetric
parameters $\delta_{H;S}$ and $\Delta_{H;S}$  from chiral breaking terms.

\section{Prediction for the Spectrum of Odd- and Even-Parity Bottom Mesons}
Using the results from charm mesons, one can
predict the spectra of the $B$ mesons.
To this end, the hyperfine operators in the theory, i.e., the parameters $\xi_{H;S}$, $T_{H;S}$
that break heavy quark symmetry,
will be rescaled to
define the mass formula for the odd- and even-parity bottom mesons.
The rescaling
can be achieved
 by  multiplying these operators by 
the ratio of the finite charm and bottom quark masses, $\frac{m_c}{m_b}$. 

The masses of the charm and bottom quarks are not directly measured. 
Many theoretical and computational methods have been developed to extract their values; for a review, see Refs. \cite{pdg12,quarkmass}.
In Table \ref{table:4.2*}, we list
  the charm and bottom quark masses evaluated from different mass schemes.
Clearly, the extracted masses of the charm and bottom quarks are not uniquely defined.
The values depend on the definition of the mass scheme
used. It is not clear which is the best definition for our purposes.
However, as the
$\overline{\mathrm{MS}}$ definition has a small associated uncertainty,
it is convenient to choose the ratio obtained from
it 
and add an extra uncertainty, of the order  $O(\Lambda_{\mathrm{QCD}})$, to cover the spread of  $\frac{m_c}{m_b}$ resulting from
different mass schemes.
Thus, the hyperfine operators in our theory can be rescaled by the factor $\frac{m_c}{m_b}=0.305\pm0.05$.
\begin{table}[t!]
\def\arraystretch{1.5}
\begin{center}
\begin{tabular}{|c|c|c|c|}
\hline
$\mathrm{Mass\,Scheme}$&$\mathrm{Charm\, quark\, mass\,(GeV)}$&$\mathrm{Bottom\, quark\, mass\,(GeV)}$&$\frac{m_c}{m_b}$\\ \hline\hline
$\overline{\mathrm{MS}}$ \cite{pdg12}& $1.275\pm0.025$&$4.18\pm0.03$&$0.305$\\  \hline
$\mathrm{Pole}$ \cite{pdg12}& $1.67\pm0.07$&$4.78\pm0.06$&$0.349$\\  \hline
$\mathrm{1\,S}$ \cite{pdg12}&....&$4.66\pm0.03$&$..$\\  \hline
$\mathrm{Kinetic}$ \cite{quarkmasstable}&$1.077\pm 0.074$&$4.549\pm0.049$&$0.237$\\  \hline
\end{tabular}
\caption{The charm and bottom $\overline{\mathrm{MS}}$ masses
are evaluated at their own scale, i.e., $\overline{m}_c(\overline{m}_c)$ and
$\overline{m}_b(\overline{m}_b)$. In Ref. \cite{pdg12}, the $\overline{\mathrm{MS}}$ values
are converted to the pole scheme. The ratio of charm and bottom masses obtained from
the pole mass is close to the ratio of the pseudoscalar charm and bottom mesons $\frac{m_D}{m_B}=0.35$.
In the kinetic mass scheme, the charm and bottom masses
are evaluated at $\mu=1\,\mathrm{GeV}$ \cite{quarkmasstable}.}
\label{table:4.2*}
\end{center}
\end{table}
In terms of  the rescaled parameters, the mass formulas for the bottom mesons up
to  one-loop corrections  are
\begin{equation}
\begin{split}
&m_{B_a}=\eta_H-\frac{3}{4}\frac{m_c}{m_b}\xi_H+\frac{\alpha_a}{3} L_{H}+\frac{\beta_a}{2} \frac{m_c}{m_b}T_{H}+\Sigma_{B_a},\\[2ex]
&m_{B^*_a}=\eta_H+\frac{1}{4}\frac{m_c}{m_b}\xi_H+\frac{\alpha_a}{3} L_{H}+\frac{\beta^*_a}{2} \frac{m_c}{m_b}T_{H}+\Sigma_{B^*_a},\\[2ex]
&m_{B_{a0}}=\eta_S-\frac{3}{4}\frac{m_c}{m_b}\xi_S+\frac{\alpha_a}{3} L_{S}+\frac{\beta_a}{2}\frac{m_c}{m_b} T_{S}+\Sigma_{B_{a0}},\\[2ex]
&m_{B^*_{a0}}=\eta_S+\frac{1}{4}\frac{m_c}{m_b}\xi_S+\frac{\alpha_a}{3} L_{S}+\frac{\beta^*_a}{2}\frac{m_c}{m_b} T_{S}+\Sigma_{B^*_{a0}},
\end{split}
\end{equation}
where the self-energy $\Sigma_B$ is a function of
the mass difference of the $B$  mesons 
and the masses of the light pseudoscalar mesons $\pi,\, \eta$, and $K$.

\begin{figure}
\centering
\begin{minipage}[c]{0.45\textwidth}
\subfloat[ ]{\includegraphics[width=\linewidth]{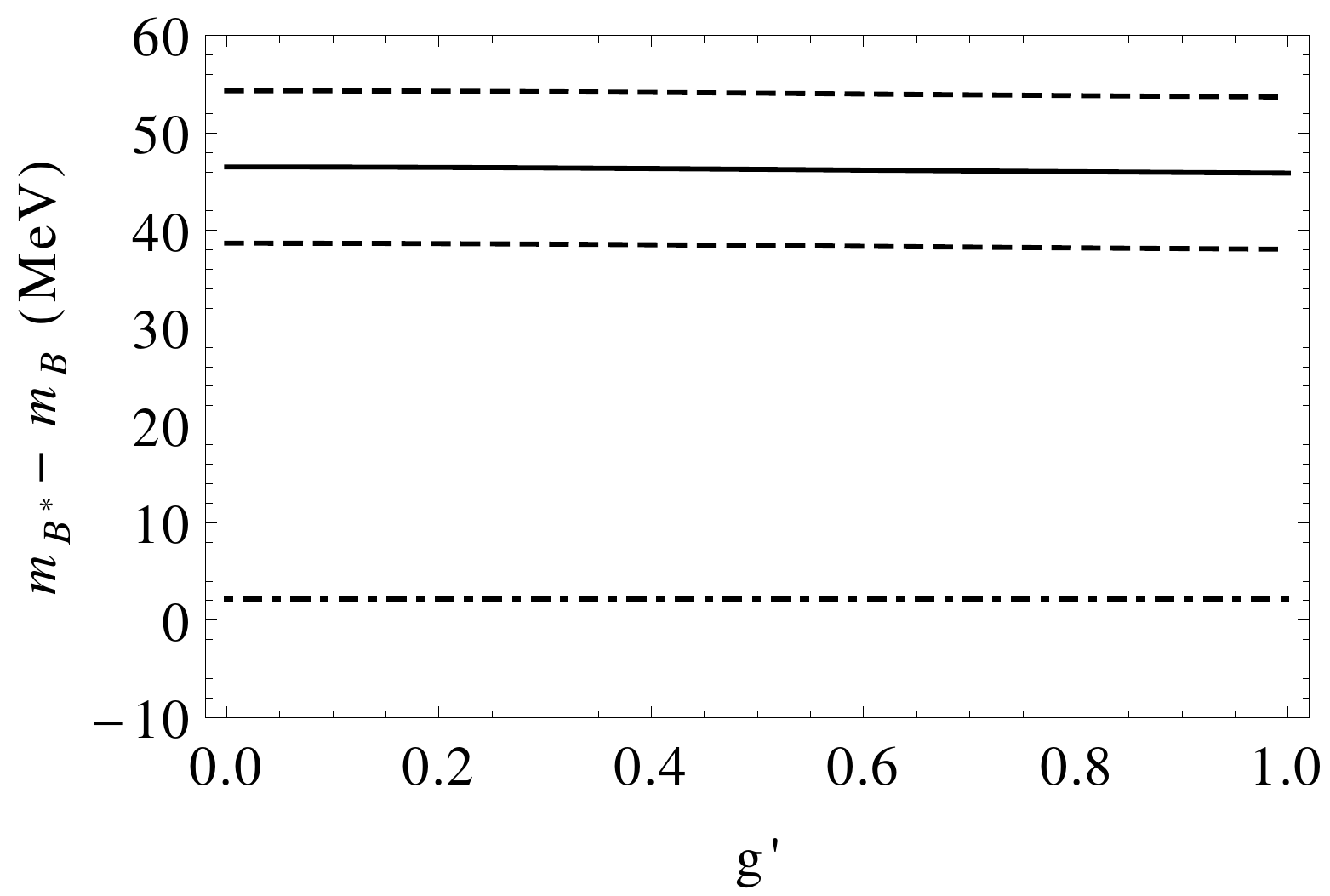}}
\end{minipage}
\vspace*{0.5cm}
\begin{minipage}[c]{0.45\textwidth}
\subfloat[]{\includegraphics[width=\linewidth]{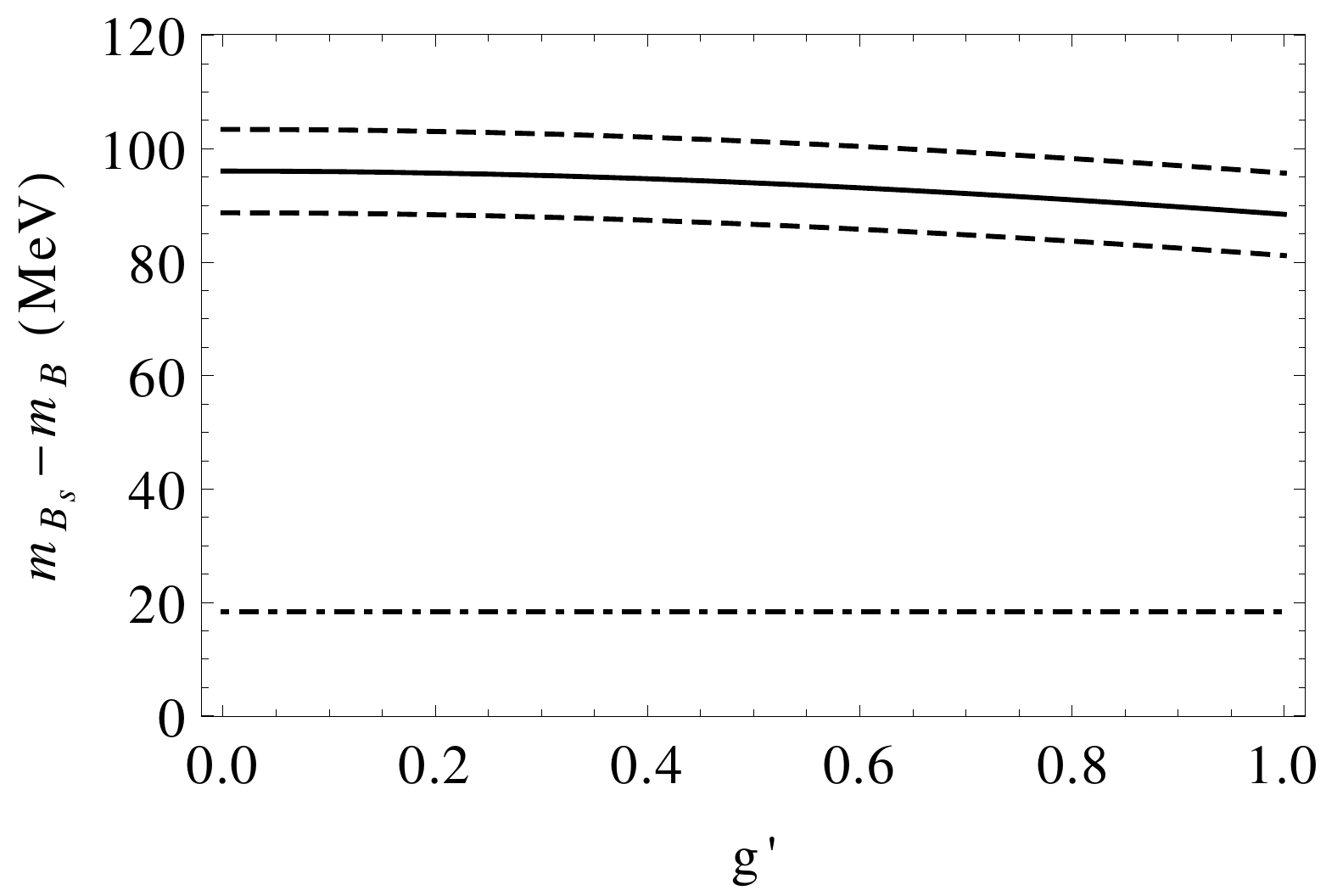}}
\end{minipage}
\begin{minipage}[c]{0.45\textwidth}
\subfloat[]{\includegraphics[width=\linewidth]{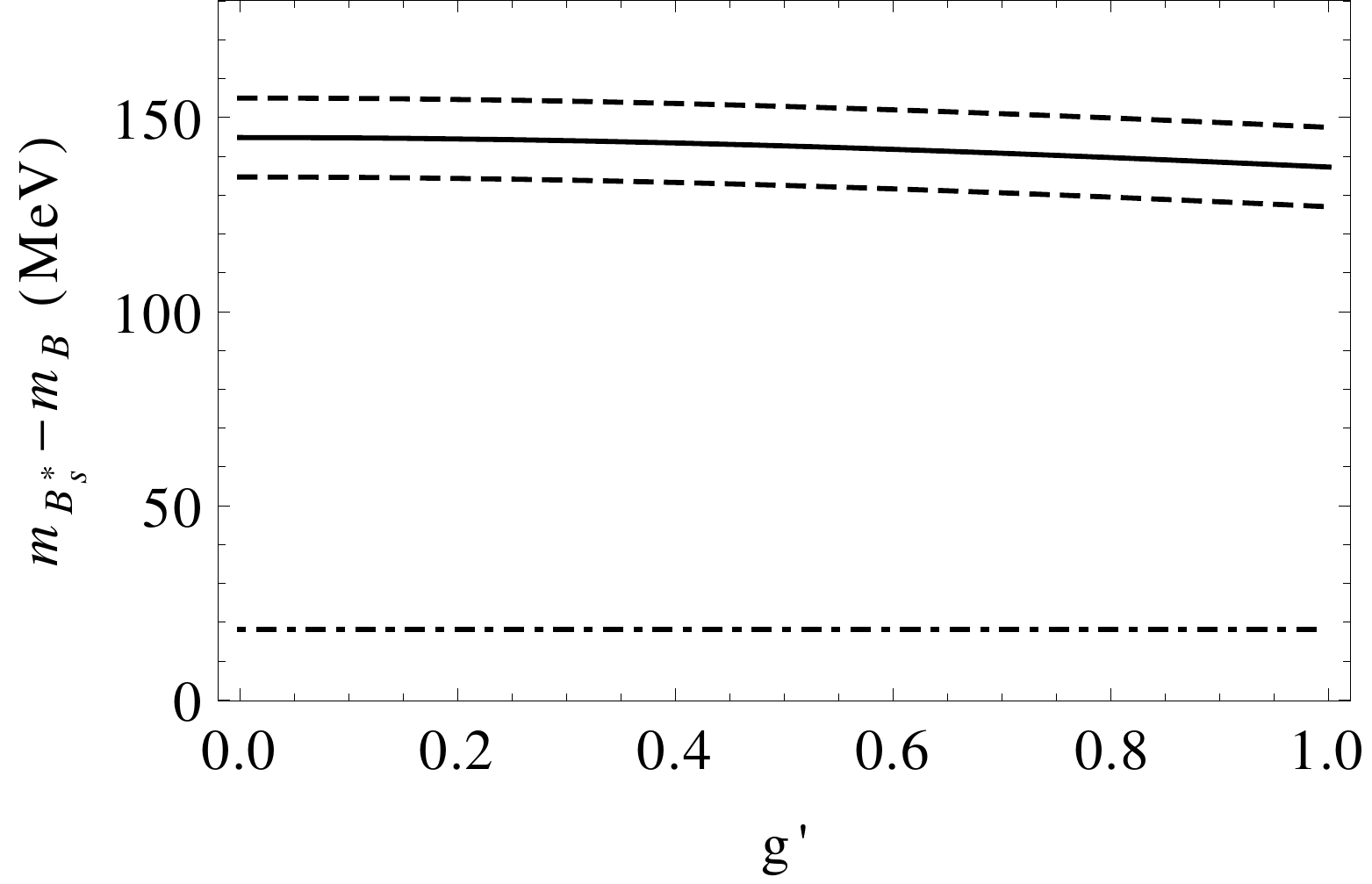}}
\end{minipage}
\vspace*{0.45cm}
\begin{minipage}[c]{0.45\textwidth}
\centering
\caption{The mass splittings plotted against $g^\prime$:  (a) $m_{B^*}-m_B$, (b) $m_{B_s}-m_B$, and (c) $m_{B_s^*}-m_B$.
The solid line represents the central value of the splittings. The associated uncertainties,
which include the experimental errors of the charm meson masses and the coupling
constants  and the error from the input parameter $\frac{m_c}{m_b}$, are given by the dashed lines.
The dotted-dashed line represents an estimate theoretical
uncertainty.}
\label{figOdd}
\end{minipage}
\end{figure}

\begin{figure}
\subfloat[]{\includegraphics[width = 0.5\textwidth]{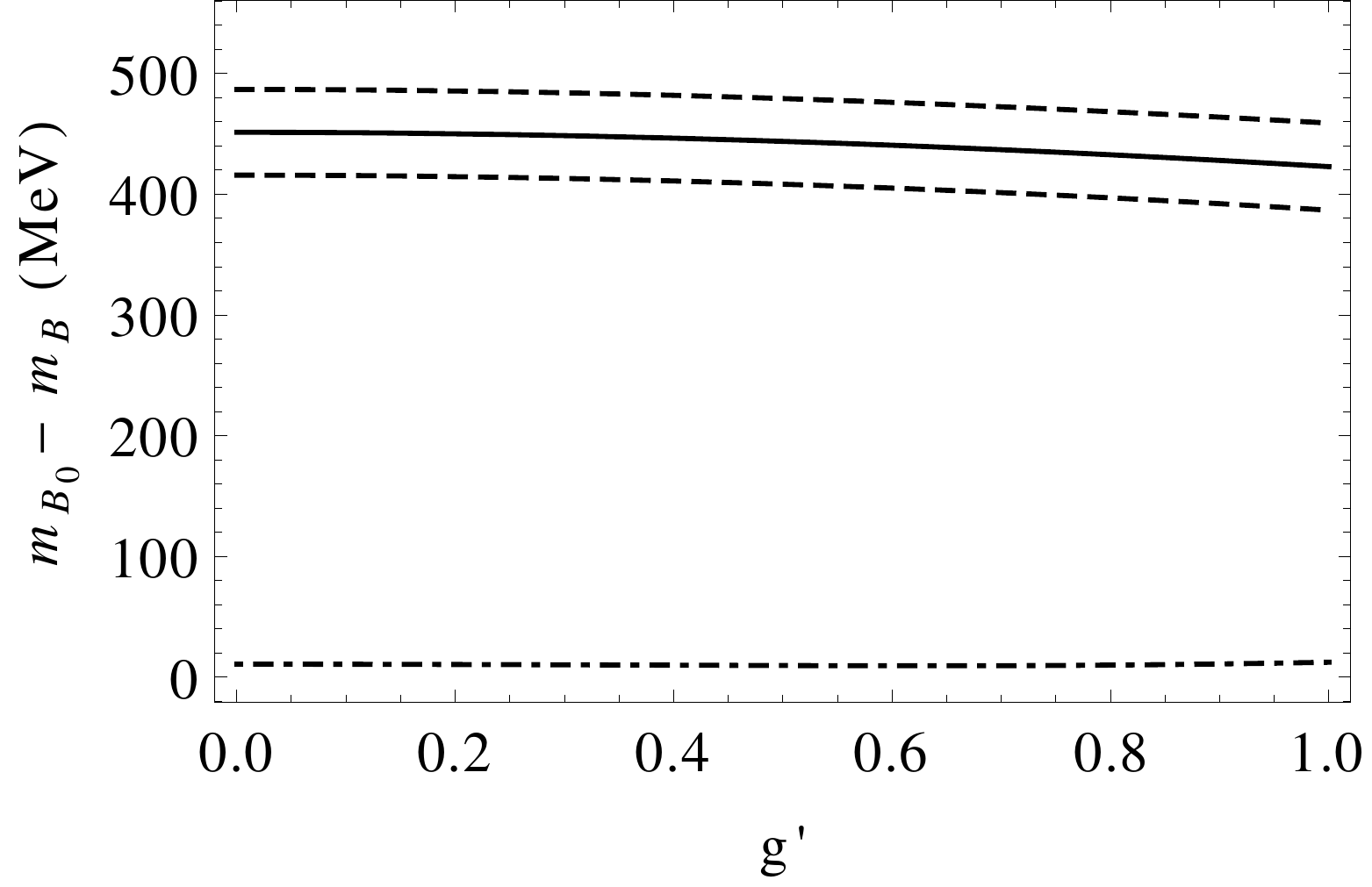}} 
\subfloat[]{\includegraphics[width = 0.5\textwidth]{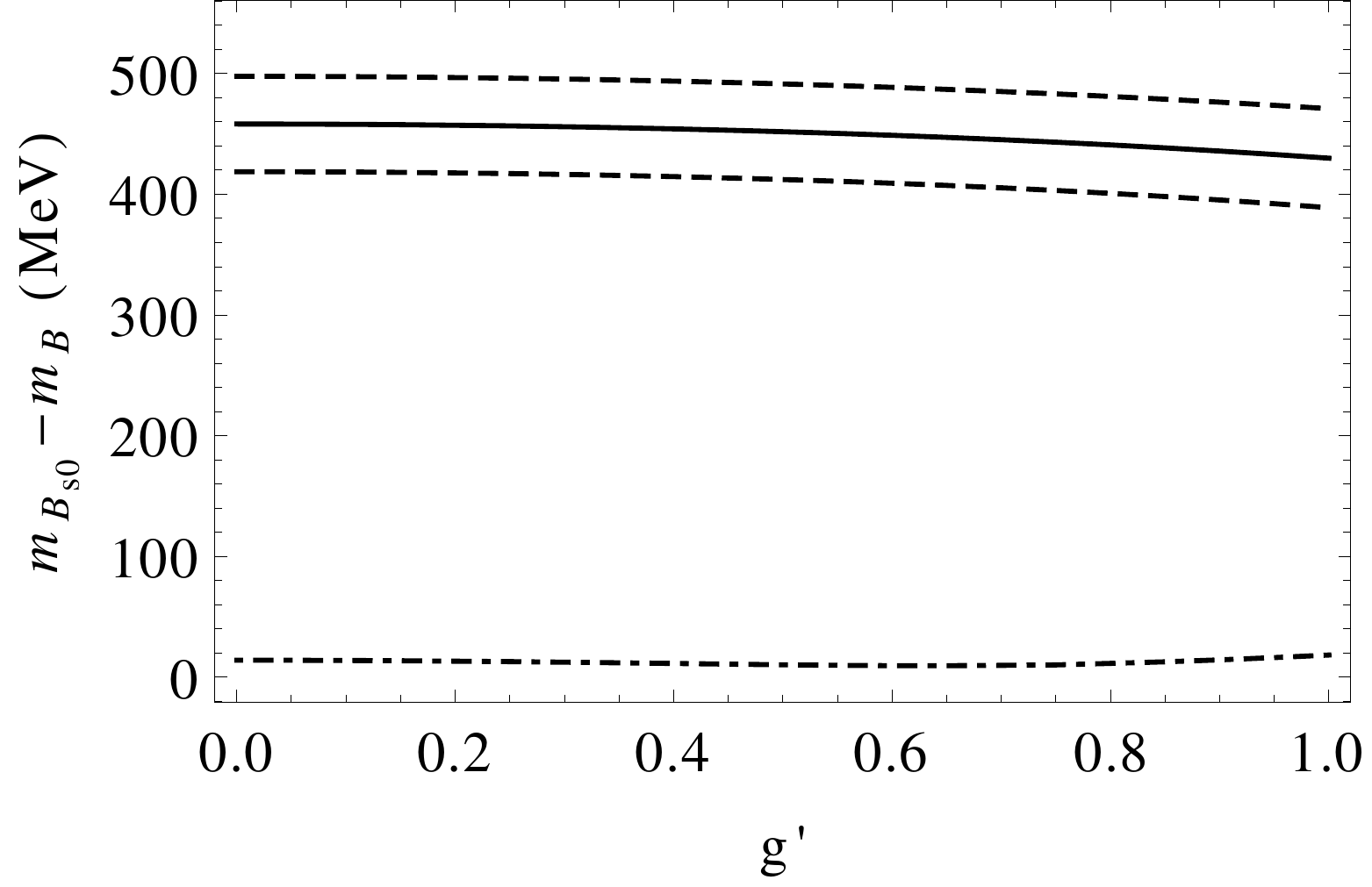}}\\ 
\subfloat[]{\includegraphics[width = 0.5\textwidth]{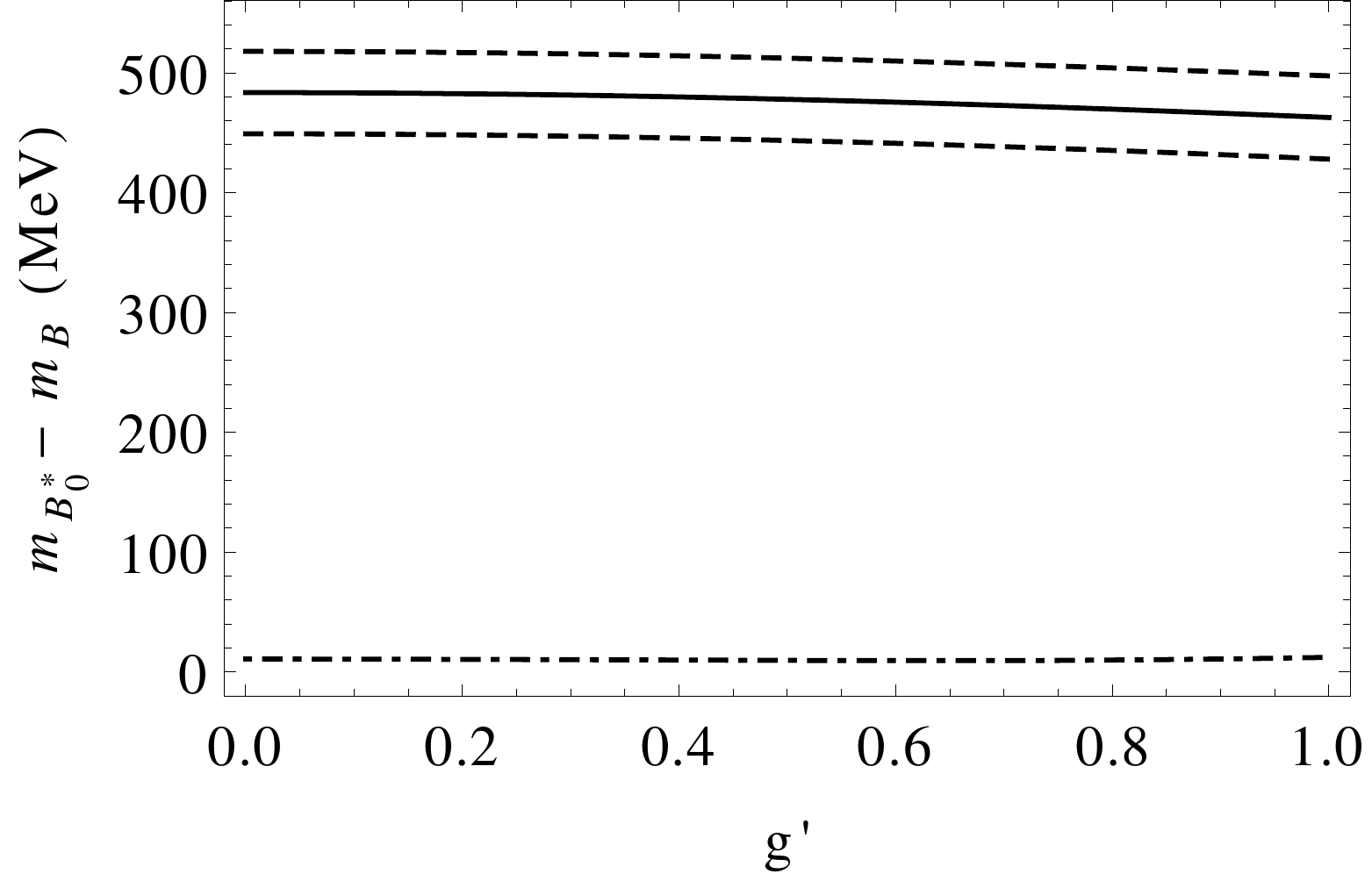}}
\subfloat[]{\includegraphics[width = 0.5\textwidth]{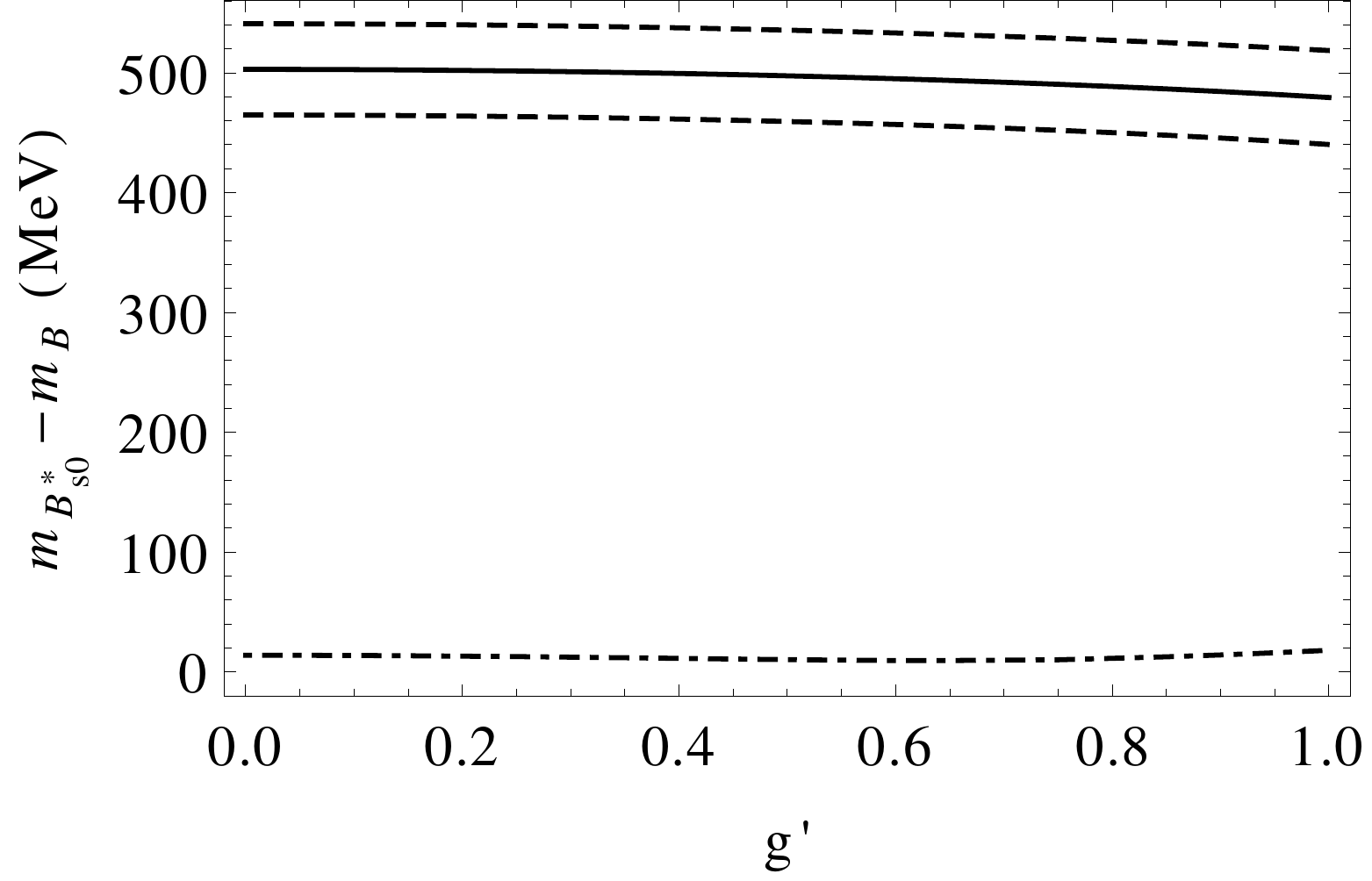}}
\caption{The mass splittings plotted against $g^\prime$:  (a) $m_{B_0}-m_B$, (b) $m_{B_{s0}}-m_B$,  (c) $m_{B_0^*}-m_B$, and (d) $m_{B_{s0}^*}-m_B$.
The notation is the same as in Fig.~\ref{figOdd}.}
\label{figEven}
\end{figure}

To predict the masses of the bottom mesons, it is suitable to choose
the ground state of the nonstrange B meson
as the
reference mass to get the following independent splittings $m_{B^*}-m_{B}$, $m_{B_s}-m_{B}$, $m_{B^*_s}-m_{B}$,
$m_{B_0}-m_{B}$, $m_{B_{s0}}-m_{B}$, $m_{B^*_0}-m_{B}$, and $m_{B^*_{s0}}-m_{B}$
where the symbols
$B$, $B_s$, $B^*$, $B^*_s$, $B_0$, $B_{s0}$, $B^*_0$, $B^*_{s0}$
represent the nonstrange pseudoscalar, strange pseudoscalar, nonstrange vector,
strange vector, nonstrange scalar, strange scalar, nonstrange axial-vector and strange axial-vector, respectively.
The loop functions depend on
the  mass differences, and so
these independent splittings form nonlinear equations.
We have used an iterative method to solve them
starting from the tree-level masses. 
The numerical values of  these mass splittings are
shown in Figs.~\ref{figOdd} and ~\ref{figEven}.

Our theoretical prediction for masses (splittings) of the odd-parity $B$ mesons are in good agreement with
the available experimental data.
In the $\mathrm{PDG}$ \cite{pdg12}, the splittings within odd-parity $B$ mesons are
\begin{eqnarray}
\label{msplitting0}
m_{B^*}-m_{B}=45.38\pm0.30~\mathrm{MeV},\\[2ex]\label{msplitting1}
m_{B^{*+}}-m_{B^+}=45.0\pm0.4~\mathrm{MeV},\\[2ex]\label{msplitting2}
m_{B_s}-m_{B}=87.33\pm0.23~\mathrm{MeV},\\[2ex]
m_{B_s^*}-m_{B_s}=48.6\pm2.41~\mathrm{MeV}.
\end{eqnarray}
The  mass difference $m_{B_s^*}-m_{B}$ can be obtained from the above splittings as follows:
\begin{align}\label{msplitting3}\nonumber
m_{B_s^*}-m_{B}&=(m_{B_s^*}-m_{B_s})+(m_{B_s}-m_{B})\\
&=135.93\pm2.42~\mathrm{MeV}.
\end{align}
 
By comparing the results in Eq.~\eqref{msplitting0} and Eq.~\eqref{msplitting1} with the predicted splitting
shown in Fig.~\ref{figOdd}(a), we find that the
experimental measurement of hyperfine splitting of the nonstrange $B$ mesons
agrees with our theoretical prediction within $1\sigma$ standard deviation.
Similarly, the measured mass difference $m_{B_s}-m_{B}$ [see Eq.~\eqref{msplitting2}]  agrees
with our theoretical prediction [see Fig.~\ref{figOdd}(b)] within about $1\sigma$ standard deviation.
Furthermore, the measured mass difference $m_{B_s^*}-m_{B}$ [see Eq.~\eqref{msplitting3}]  agrees
with our theoretical prediction [see Fig.~\ref{figOdd}(c)]  within $1\sigma$ standard deviation.

For the even-parity sector, the $B$-meson states have not yet been observed; thus,
our results, which are shown in Fig.~\ref{figEven}, provide useful
information for  experimentalists  investigating such states.

For the predicted masses (splittings) of the even-parity sector,
the strong dependence on the coupling $g^{\prime}$ is
due to the large negative contribution from
 terms with
 \begin{eqnarray}\nonumber
\frac{g^{\prime 2}}{4\,f^2}\,n_f\,K_1(\omega,m)& \simeq&  \frac{g^{\prime 2}}{4\,f^2}\,n_f\,\Big(- \frac{4}{16\,\pi^2}(\omega^2-m^2)~F(\omega,m)+...\Big)\propto -\frac{g^{\prime 2}}{f^2}\,n_f\, m^2 \sqrt{m^2-\omega^2} \cos^{-1}\left(\frac{\omega}{m}\right)+...,
\end{eqnarray}
for $m^2>\omega^2$ where $m=m_{\eta},\,m_K$. The light-quark
factor  $n_f$ is simply obtained from the Gell-Mann matrices,
and its value reflects the number of independent self-energy loop
 diagrams which contribute to the process.
 
Before proceeding to comment on the $SU(3)$-splittings within the predicted $B$-meson states, let us first briefly examine 
the charm meson masses given in
Table \ref{table:1.4bb}. Evidently, the strange and nonstrange  splittings of the well
determined states, i.e.,  $J^{p}=0^-$ and $J^{p}=1^-$,
are consistent with the size of the $SU(3)$ breaking, $O(100~ \mathrm{MeV})$. 
However, this is not the case for 
the even-parity sector where the central values of the splittings
\begin{equation}\label{Dsplitting}
 m_{D^{*\pm}_{s0}}-m_{D^{*0}_0}= -0.3\pm29~\mathrm{MeV},~~m_{D^{\pm\prime}_{s1}}-m_{D^{0\prime}_0}= 32.5\pm36~\mathrm{MeV},
\end{equation}
are inconsistent with the size of the $SU(3)$ breaking.
The closeness of $m_{D^{*\pm}_{s0}}$ and $m_{D^{*0}_0}$ masses
was the first observation of
mass degeneracy in heavy-light mesons.

From the heavy quark symmetry, the observed mass degeneracy in the charm sector
implies the similarity of $m_{B_0}$ and $m_{B_{s0}}$ in the bottom sector.
Our approach of using HHChPT shows that
there is an accidental cancellation between   
$SU(3)$-breaking loop contributions and counterterms in the even-parity $B$-meson sector.
Hence, it is obvious from Figs.~\ref{figEven}(a) and \ref{figEven}(b) that the nonstrange and strange scalar bottom mesons are nearly degenerate, 
the difference between their central
values is $\sim 8\,\mathrm{MeV}$. Moreover,
the splitting between nonstrange and strange axial-vector 
bottom mesons is $\sim 19\,\mathrm{MeV}$; see Figs.~\ref{figEven}(c) and \ref{figEven}(d). 
This result, which is inconsistent with the theoretical expectation on $SU(3)$ breaking, was observed in charm sector.

The difference between 
$SU(3)$-splittings in the predicted $B$-meson sector is approximately equal to the ones in the observed $D$-meson sector
 times the rescaling factor, i.e.,  
\begin{equation}
(m_{B^*_{s0}}-m_{B^*_0})-(m_{B_{s0}}-m_{B_0})\approx \frac{m_c}{m_b} \left[(m_{D^{\pm\prime}_{s1}}-m_{D^{0\prime}_0})-(m_{D^{*\pm}_{s0}}-m_{D^{*0}_0})\right].
\end{equation}
This mass relation is consistent with the heavy quark spin-flavor symmetry.

It is worth mentioning that
the work undertaken in Refs. \cite{GKM,bmeson} was intended
to investigate the closeness of nonstrange and strange scalars in
the charm and bottom
sectors, using, in addition to $\mathrm{HHChPT}$, different potential models.
They considered the hadronic loops effect to shift
down the bare masses of scalar mesons. 
In their work, the hadronic loop contributions include only
 the coupling of $D^*_{s0}$ to the lowest possible intermediate states,
these states form members of $\frac{1}{2}^-$-doublet
in the notation of $\mathrm{HHChPT}$.
The self-energy contributions from the coupling of
$D^*_{s0}$ to the members of the $\frac{1}{2}^+$-doublet
have been neglected in Refs. \cite{GKM,bmeson}
which in turn indicates their analysis within $\mathrm{HHChPT}$ is incomplete.
In Ref. \cite{bmeson}, the authors 
concluded that the results
of studying  the mass degeneracy using $\mathrm{HHChPT}$ are
not satisfactory, which is in fact not true as shown in Figs.~\ref{figEven}(a) and \ref{figEven}(b).

Furthermore, the approach employed in Refs. \cite{GKM,bmeson}
of using bare masses in evaluating loop functions
is
inappropriate for the case of $\mathrm{HHChPT}$.
For example, the
 predicted masses of $B^*_{s0}$ and $B^*_0$, as given
in TABLE II in Ref. \cite{bmeson},
provide different
splittings when using different bare masses in
evaluating loop functions.
More precisely,
the mass difference $m_{B^*_{s0}}-m_{B^*_0}$ is $\sim + 100\,\mathrm{MeV}$
when  evaluating loop functions
with bare masses given in Ref. \cite{bare1} and  is $\sim -60\,\mathrm{MeV}$  when  evaluating loop functions
 with bare masses taken from Ref. \cite{bare2}.
This shows that the loop integrals are sensitive to the input mass
differences of the heavy mesons,
so 
using bare masses 
is not appropriate.
To avoid
these problems, we use the self-consistently determined  masses. 
As a result, there is an unavoidable theoretical uncertainty, which we estimate 
from  higher-order contributions from the $\beta$ function. 

\section{Summary}
The  aspects  of
 mesons containing a single heavy quark are governed by the spin symmetry $SU(2)_s$ of the heavy quark and the chiral symmetry $SU(3)_L\times SU(3)_R$ of the light quarks.
Incorporating both approximate symmetries in a single framework was achieved by defining the heavy hadron chiral perturbation theory. 
This effective theory was used
to study the spectra and interactions of these heavy mesons.
We studied the masses of the low-lying charm and bottom mesons
using $\mathrm{HHChPT}$.
We expressed the masses of these heavy mesons
 up to third order, $Q^3$, in the chiral expansion,
where meson loops contribute. 
The heavy-hadron chiral Lagrangian has $12$  unknown low-energy constants 
($\delta_{H;S}$, $a_{H;S}$, $\sigma_{H;S}$, $\Delta_{H;S}$, $\Delta^{(a)}_{H;S}$,  $\Delta^{(\sigma)}_{H;S}$)
to describe eight measured masses of charm mesons. Hence,
obtaining unique numerical values of the LECs is impossible.
We used flavor and heavy quark symmetries 
to construct eight linear combinations
($\eta_{H;S}$, $\xi_{H;S}$, $L_{H;S}$, $T_{H;S}$)
out of  the LECs. 
By using this method, we reduced the
number of unknown LECs to be comparable with the current experimental data on
meson masses.
Thus, 
 one can express these parameters directly in terms of
the physical masses and loop integrals.
In contrast to previous approaches, we used physical meson masses in evaluating
the heavy meson loops.
As a result, the energy of any unstable particle is placed correctly
 relative to the decay threshold, and the imaginary part of the loop integral
can be related to the experimental decay width.
However, the resulting values for these parameters
 contain contributions beyond the order $Q^3$ of heavy-hadron chiral Lagrangian.
 This is due to using empirical masses which
generate higher order  $\mu$-dependent terms that
 cannot be renormalized  using $\mu$-dependent
counterterms of our Lagrangian. 
To this end, we chose to define the $\beta$ functions for these parameters
to estimate the contributions from higher-order terms.
Having  fitted the linear combinations of the LECs to the $D$-meson spectrum, we rescale
the hyperfine combinations
to predict  the masses of odd- and even-parity bottom mesons.
In our calculations, we used a self-consistent approach to extract the $B$-meson masses;
i.e., the values we started with to evaluate the mass splittings within $B$-meson states are
the same as the resultant mass splittings. 
The predicted masses from our theory are
in good agreement with experimentally measured masses
for the case of the odd-parity sector.
For the even-parity sector, the $B$-meson states have not yet been observed; thus,
our results provide useful
information for  experimentalists  investigating such states.

The approach developed in this paper
can be extended to
predict the spectra of the other doublet of the $P$-wave states, i.e., $S^p=\frac{3}{2}^+$,
where $S$ is the total angular momentum of the light degrees of freedom, and $p$ is the parity.
The spin-parity quantum numbers of
these states are $1^+$ and $2^+$. 
This requires introducing a new (tensor) field to describe the dynamics of these states
in the  chiral
Lagrangian. 
The general structure of the relevant chiral
Lagrangian with tensor fields is represented in \cite{cdgn12,cas97,FalkLuke} for instance.
\begin{acknowledgments}
I am grateful to Michael C. Birse, whose guidance and support helped
me to develop an understanding of the subject.
\end{acknowledgments}

\appendix
\section{Self-Energies of Charm Mesons}
The explicit expressions for  the self-energies of the charm mesons  are
\begin{equation}
\begin{split}
 \Sigma_{H_1}&=\frac{g^2}{4 f^2}\left[ 3K_1(m_{H^*_1}-m_{H_1},m_{\pi})+\frac{1}{3}K_1(m_{H^*_1}-m_{H_1},m_{\eta})+2 K_1(m_{H^*_3}-m_{H_1},m_K) \right]\\[2ex]
&+\frac{h^2}{4 f^2}\left[ 3K_2(m_{S_1}-m_{H_1},m_{\pi})+\frac{1}{3}K_2(m_{S_1}-m_{H_1},m_{\eta})+2 K_2(m_{S_3}-m_{H_1},m_K) \right],\\
\end{split}
\end{equation}
\begin{equation}
\begin{split}
 \Sigma_{H^*_1}&=\frac{g^2}{4 f^2}\left[ K_1(m_{H_1}-m_{H^*_1},m_{\pi})+\frac{1}{9}K_1(m_{H_1}-m_{H^*_1},m_{\eta})+\frac{2}{3} K_1(m_{H_3}-m_{H^*_1},m_K) \right]\\[2ex]
&+\frac{g^2}{4 f^2}\left[ 2K_1(0,m_{\pi})+\frac{2}{9}K_1(0,m_{\eta})+\frac{4}{3} K_1(m_{H^*_3}-m_{H^*_1},m_K) \right]\\[2ex]
&+\frac{h^2}{4 f^2}\left[ 3K_2(m_{S^*_1}-m_{H^*_1},m_{\pi})+\frac{1}{3}K_2(m_{S^*_1}-m_{H^*_1},m_{\eta})+2 K_2(m_{S^*_3}-m_{H^*_1},m_K) \right],\\
\end{split}
\end{equation}
\begin{equation}
\begin{split}
\Sigma_{H_3}&=\frac{g^2}{4 f^2}\left[\frac{4}{3}K_1(m_{H^*_3}-m_{H_3},m_{\eta})+4 K_1(m_{H^*_1}-m_{H_3},m_K) \right]\\[2ex]
&+\frac{h^2}{4 f^2}\left[ \frac{4}{3}K_2(m_{S_3}-m_{H_3},m_{\eta})+4 K_2(m_{S_1}-m_{H_3},m_K) \right],\\
\end{split}
\end{equation}

\begin{equation}
\begin{split}
\Sigma_{H^*_3}&=\frac{g^2}{4 f^2}\left[ \frac{4}{9}K_1(m_{H_3}-m_{H^*_3},m_{\eta})+\frac{4}{3} K_1(m_{H_1}-m_{H^*_3},m_K) \right]\\[2ex]
&+\frac{g^2}{4 f^2}\left[\frac{8}{9}K_1(0,m_{\eta})+\frac{8}{3} K_1(m_{H^*_1}-m_{H^*_3},m_K) \right]\\[2ex]
&+\frac{h^2}{4 f^2}\left[\frac{4}{3}K_2(m_{S^*_3}-m_{H^*_3},m_{\eta})+4 K_2(m_{S^*_1}-m_{H^*_3},m_K) \right],\\
\end{split}
\end{equation}
\begin{equation}
\begin{split}
 \Sigma_{S_1}&=\frac{g^{\prime 2}}{4 f^2}\left[ 3K_1(m_{S^*_1}-m_{S_1},m_{\pi})+\frac{1}{3}K_1(m_{S^*_1}-m_{S_1},m_{\eta})+2 K_1(m_{S^*_3}-m_{S_1},m_K) \right]\\[2ex]
&+\frac{h^2}{4 f^2}\left[ 3K_2(m_{H_1}-m_{S_1},m_{\pi})+\frac{1}{3}K_2(m_{H_1}-m_{S_1},m_{\eta})+2 K_2(m_{H_3}-m_{S_1},m_K) \right],\\
\end{split}
\end{equation}
\begin{equation}
\begin{split}
 \Sigma_{S^*_1}&=\frac{g^{\prime 2}}{4 f^2}\left[K_1(m_{S_1}-m_{S^*_1},m_{\pi})+\frac{1}{9}K_1(m_{S_1}-m_{S^*_1},m_{\eta})+\frac{2}{3} K_1(m_{S_3}-m_{S^*_1},m_K) \right]\\[2ex]
&+\frac{g^{\prime 2}}{4 f^2}\left[2K_1(0,m_{\pi})+\frac{2}{9}K_1(0,m_{\eta})+\frac{4}{3} K_1(m_{S^*_3}-m_{S^*_1},m_K) \right]\\[2ex]
&+\frac{h^2}{4 f^2}\left[3K_2(m_{H^*_1}-m_{S^*_1},m_{\pi})+\frac{1}{3}K_2(m_{H^*_1}-m_{S^*_1},m_{\eta})+2 K_2(m_{H^*_3}-m_{S^*_1},m_K) \right],\\
\end{split}
\end{equation}
\begin{equation}
\begin{split}
\Sigma_{S_3}&=\frac{g^{\prime 2}}{4 f^2}\left[\frac{4}{3}K_1(m_{S^*_3}-m_{S_3},m_{\eta})+4 K_1(m_{S^*_1}-m_{S_3},m_K) \right]\\[2ex]
&+\frac{h^2}{4 f^2}\left[\frac{4}{3}K_2(m_{H_3}-m_{S_3},m_{\eta})+4 K_2(m_{H_1}-m_{S_3},m_K) \right],\\
\end{split}
\end{equation}

\begin{equation}
\begin{split}
\Sigma_{S^*_3}&=\frac{g^{\prime 2}}{4 f^2}\left[\frac{4}{9}K_1(m_{S_3}-m_{S^*_3},m_{\eta})+\frac{4}{3} K_1(m_{S_1}-m_{S^*_3},m_K) \right]\\[2ex]
&+\frac{g^{\prime 2}}{4 f^2}\left[\frac{8}{9}K_1(0,m_{\eta})+\frac{8}{3} K_1(m_{S^*_1}-m_{S^*_3},m_K) \right]\\[2ex]
&+\frac{h^2}{4 f^2}\left[\frac{4}{3}K_2(m_{H^*_3}-m_{S^*_3},m_{\eta})+4 K_2(m_{H^*_1}-m_{S^*_3},m_K) \right].
\end{split}
\end{equation}
The chiral loop integrals are
\begin{equation}
\begin{split}
K_1(\omega, m)&=\frac{1}{16 \pi^2}\left[(-2\omega^3+3m^2\omega)\mathrm{ln}\left(\frac{m^2}{\mu^2}\right)-4(\omega^2- m^2)F(\omega,m)+\frac{16}{3}\omega^3-7\omega\, m^2\right],\\[2ex]
K_2(\omega, m)&=\frac{1}{16 \pi^2} \left[ (-2\omega^3+ m^2\omega)\mathrm{ln}\left(\frac{m^2}{\mu^2}\right)-4\omega^2 F(\omega,m)+4\omega^3- \omega\, m^2\right],
\end{split}
\end{equation}
 renormalized in the $\mathrm{\overline{MS}}$ scheme. 
  The function $F(\omega,m)$ is given by 
\begin{equation}\label{F-function}
 F(\omega,m)=\left\{ \begin{array}{c c}-\sqrt{m^2-\omega^2} \cos^{-1}(\frac{\omega}{m}), & \mbox{$m^2>\omega^2,$} \\[2ex]
                                 \sqrt{\omega^2-m^2}[i \pi-\cosh^{-1}(-\frac{\omega}{m})], & \mbox{$\omega<-m,$}\\[2ex]
 \sqrt{\omega^2-m^2}\cosh^{-1}(\frac{\omega}{m}), & \mbox{$\omega>m.$}
                                \end{array} \right.
\end{equation}
It is worth mentioning that
the expression for $K_1(\omega, m)$
in Ref. \cite{ms05} does not agree with our expression.
Some finite pieces are missed  due to the inconsistent use of
 dimensional regularization; i.e.,
the authors set $d=4$ before expanding in powers of $4-d$.
However, our calculation when using the chiral function $K_1(\omega, m)$
from Ref. \cite{ms05}, i.e.,  
\begin{equation}\label{ms05}
K_1(\omega, m)=\frac{1}{16 \pi^2}\left[(-2\omega^3+3m^2\omega)\mathrm{ln}\left(\frac{m^2}{\mu^2}\right)-4(\omega^2- m^2)F(\omega,m)+4\omega^3-5\omega\, m^2\right],
\end{equation}
does not affect much the results on the $B$ meson spectra. The difference between the obtained results using our expression and the ones in
Ref. \cite{ms05} is less than $1~\mathrm{MeV}$. However, the values of parameters that break flavor and/or spin symmetries, i.e., $\xi_{H;S}$, $L_{H;S}$, $T_{H;S}$, are much affected. 
For instance, the central values of odd-parity
parameters given in Eq.~\eqref{Odd} become
 \begin{equation}
\begin{split}
\eta_H=171.57~\mathrm{MeV},~~\xi_H=173.04~\mathrm{MeV},~~L_{H}=263.13~\mathrm{MeV},~~T_{H}=-29.54~\mathrm{MeV}.
\end{split}
\end{equation}
Our expression for $K_2(\omega, m)$ agrees with the expression presented in Ref. \cite{ms05}; for details,
see Appendix B.

\section{Calculation of Loop Corrections}
For the sake of simplicity, we restrict our discussion to $SU(2)$ HHChPT
with nonstrange $D$ mesons. Our calculations of loop diagrams differ from those in Refs. \cite{ms05,cas97,Grins}
in two aspects:\\
i) Dimensional regularization is used consistently. \\
ii) To maintain the heavy quark symmetry at the quantum loop level, the nonrelativistic heavy meson fields are defined in
four dimensions.

In Figs.~\ref{fig1} and \ref{fig2}, we show the
Feynman diagrams of the one-loop correction to the masses of $D$ mesons.
In evaluating loop integrals for these diagrams,
one has to be careful with the tensor structure to get the correct expressions. For this purpose, we will calculate
loop integrals for diagrams $a-e$ in  Fig.~\ref{fig1}. The results  hold for  diagrams with a similar tensor structure of even-parity sector as shown in Fig.~\ref{fig2}.

Let us start with the loop diagram $a$ in Fig.~\ref{fig1}, which contributes to the self-energy of the $H_1$ field, i.e., the $D^+$
\begin{equation}
\begin{split}\label{chlointps00}
 i\,\Sigma^{(a)}_{H_1}&=3\left(\frac{g}{2\,f}\right)^2\mu^{4-d} \int \frac{ d^dq}{(2\pi)^d} \frac{q^{\mu} q^{\nu} (g_{\mu \nu}-v_{\mu}v_{\nu})}{(q \cdot v-\omega_a+i \epsilon) (q^2-m^2_{\pi}+i \epsilon)}\\[2ex]
&=3\left(\frac{g}{2\,f}\right)^2(g_{\mu \nu}-v_{\mu}v_{\nu})\mu^{4-d} \int \frac{ d^dq}{(2\pi)^d} \frac{q^{\mu} q^{\nu} }{(q \cdot v-\omega_a+i \epsilon) (q^2-m^2_{\pi}+i \epsilon)},
\end{split}
\end{equation}
where $\omega$ is the mass difference between internal and external heavy meson states.
The factor $3$ results from Pauli matrices $(\tau^2_i)_{\alpha \beta}=3\,\delta_{\alpha \beta}$,
 where for one-loop diagrams in which  a single pion is exchanged $\alpha=\beta$, so $\delta_{\alpha \alpha}=1$. 

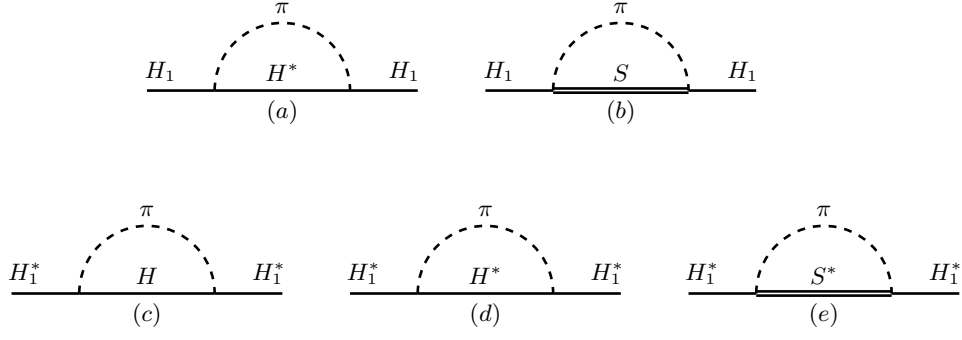
\begin{figure}[h!]
\begin{center}
\begin{tikzpicture}[domain=-5:16,scale=0.9]
\draw[-,thick,line width=1pt] (3,0)--(4,0)   (3.2,0) node[above]{$H_1$};
\draw[-,thick,line width=1pt] (4,0)--(6,0) (5,0) node[above]{$H^*$};
\node[below,line width=1pt] at   (5,0) {$(a)$};
\draw[-,thick,line width=1pt] (6,0)--(7,0)  (6.8,0) node[above]{$H_1$};
\draw [dashed,line width=1pt] (4,0) arc (180:0: 1cm) (5,1) node[above]{$\pi$};
\draw[-,thick,line width=1pt] (8,0)--(9,0)   (8.2,0) node[above]{$H_1$};
\draw[double,line width=1pt] (9,0)--(11,0) (10,0) node[above]{$S$};
\node[below,line width=1pt] at   (10,0) {$(b)$};
\draw [dashed,line width=1pt] (9,0) arc (180:0: 1cm) (10,1) node[above]{$\pi$};
\draw[-,thick,line width=1pt] (11,0)--(12,0)  (11.8,0) node[above]{$H_1$};
\draw[-,thick,line width=1pt] (1,-3)--(2,-3)   (1.2,-3) node[above]{$H^*_1$};
\draw[-,thick,line width=1pt] (2,-3)--(4,-3) (3,-3) node[above]{$H$};
\node[below,line width=1pt] at   (3,-3) {$(c)$};
\draw[-,thick,line width=1pt] (4,-3)--(5,-3)  (4.8,-3) node[above]{$H^*_1$};
\draw [dashed,line width=1pt] (2,-3) arc (180:0: 1cm) (3,-2) node[above]{$\pi$};
\draw[-,thick,line width=1pt] (6,-3)--(7,-3)   (6.2,-3) node[above]{$H^*_1$};
\draw[-,thick,line width=1pt] (7,-3)--(9,-3) (8,-3) node[above]{$H^*$};
\node[below,line width=1pt] at   (8,-3) {$(d)$};
\draw [dashed,line width=1pt] (7,-3) arc (180:0: 1cm) (8,-2) node[above]{$\pi$};
\draw[-,thick,line width=1pt] (9,-3)--(10,-3)  (9.8,-3) node[above]{$H^*_1$};
\draw[-,thick,line width=1pt] (11,-3)--(12,-3)   (11.2,-3) node[above]{$H^*_1$};
\draw[double,line width=1pt] (12,-3)--(14,-3) (13,-3) node[above]{$S^*$};
\node[below,line width=1pt] at   (13,-3) {$(e)$};
\draw [dashed,line width=1pt] (12,-3) arc (180:0: 1cm) (13,-2) node[above]{$\pi$};
\draw[-,thick,line width=1pt] (14,-3)--(15,-3)  (14.8,-3) node[above]{$H^*_1$};
\end{tikzpicture}
\caption{Feynman diagrams shown in (a) and (b) represent the self-energy of the $H_1$ field and those shown in (c)-(e) represent the self-energy of the $H^*_1$ field.}\label{fig1}
\end{center}
\end{figure}
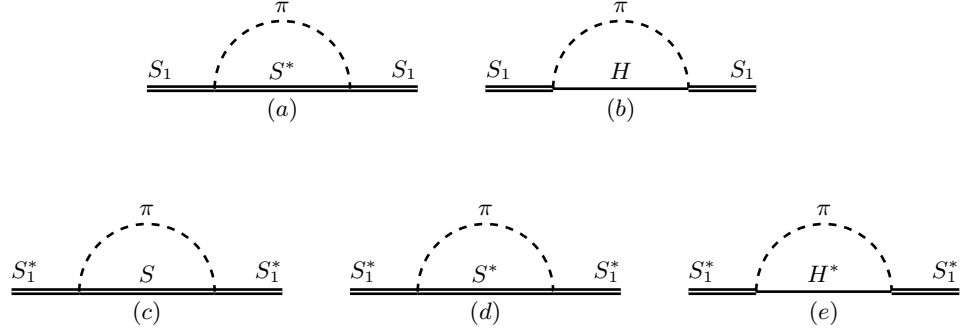
\begin{figure}[h!]
\begin{center}
\begin{tikzpicture}[domain=-5:16,scale=0.9]
\draw[double,line width=1pt] (3,0)--(4,0)   (3.2,0) node[above]{$S_1$};
\draw[double,line width=1pt] (4,0)--(6,0) (5,0) node[above]{$S^*$};
\node[below,line width=1pt] at   (5,0) {$(a)$};
\draw[double,line width=1pt] (6,0)--(7,0)  (6.8,0) node[above]{$S_1$};
\draw [dashed,line width=1pt] (4,0) arc (180:0: 1cm) (5,1) node[above]{$\pi$};
\draw[double,line width=1pt] (8,0)--(9,0)   (8.2,0) node[above]{$S_1$};
\draw[-,thick,line width=1pt] (9,0)--(11,0) (10,0) node[above]{$H$};
\node[below,line width=1pt] at   (10,0) {$(b)$};
\draw [dashed,line width=1pt] (9,0) arc (180:0: 1cm) (10,1) node[above]{$\pi$};
\draw[double,line width=1pt] (11,0)--(12,0)  (11.8,0) node[above]{$S_1$};
\draw[double,line width=1pt] (1,-3)--(2,-3)   (1.2,-3) node[above]{$S^*_1$};
\draw[double,line width=1pt] (2,-3)--(4,-3) (3,-3) node[above]{$S$};
\node[below,line width=1pt] at   (3,-3) {$(c)$};
\draw[double,line width=1pt] (4,-3)--(5,-3)  (4.8,-3) node[above]{$S^*_1$};
\draw [dashed,line width=1pt] (2,-3) arc (180:0: 1cm) (3,-2) node[above]{$\pi$};
\draw[double,line width=1pt] (6,-3)--(7,-3)   (6.2,-3) node[above]{$S^*_1$};
\draw[double,line width=1pt] (7,-3)--(9,-3) (8,-3) node[above]{$S^*$};
\node[below,line width=1pt] at   (8,-3) {$(d)$};
\draw [dashed,line width=1pt] (7,-3) arc (180:0: 1cm) (8,-2) node[above]{$\pi$};
\draw[double,line width=1pt] (9,-3)--(10,-3)  (9.8,-3) node[above]{$S^*_1$};
\draw[double,line width=1pt] (11,-3)--(12,-3)   (11.2,-3) node[above]{$S^*_1$};
\draw[-,thick,line width=1pt] (12,-3)--(14,-3) (13,-3) node[above]{$H^*$};
\node[below,line width=1pt] at   (13,-3) {$(e)$};
\draw [dashed,line width=1pt] (12,-3) arc (180:0: 1cm) (13,-2) node[above]{$\pi$};
\draw[double,line width=1pt] (14,-3)--(15,-3)  (14.8,-3) node[above]{$S^*_1$};
\end{tikzpicture}
\caption{Feynman diagrams shown in (a) and (b) represent the self-energy of the $S_1$ field and those shown in (c)-(e) represent the self-energy of the $S^*_1$ field.}
\label{fig2}
\end{center}
\end{figure}

The chiral loop integral is divergent.
However, there are many ways to regulate the above loop-integral and each one
introduces a new momentum scale of which physical observables must be independent.
In field theory, the so-called dimensional regularization scheme ($\mathrm{DR}$)
is widely used since it preserves gauge and chiral symmetries as well as
Lorentz (Galilean) invariance for relativistic (nonrelativistic) systems.

For loop integrals containing  two or more powers of $q$ (momentum of the internal pion) in the numerator,
the standard procedure of evaluating them 
is  to break them up into simple  integrals that can then be  easily  calculated \cite{bkm95,sch03}.
Thus, one can write
\begin{equation}\label{eq02012}
i \mu^{4-d} \int \frac{ d^dq}{(2\pi)^d} \frac{q^{\mu} q^{\nu} }{(q \cdot v-\omega+i \epsilon) (q^2-m^2_{\pi}+i \epsilon)}= g^{\mu \nu} J_2+v^{\mu}v^{\nu} J_3,
\end{equation}
where
\begin{equation}\label{eq39}
 J_2=\frac{1}{d-1}[(m^2_{\pi}-\omega^2)J_0-\omega J_{\pi}],
\end{equation}
and
\begin{equation}\label{eq40}
 J_3=\frac{1}{d-1}[(d\, \omega^2 -m^2_{\pi})J_0+ \omega \,d\, J_{\pi}].
\end{equation}
The explicit expression for $J_0$ is
\begin{equation}
\begin{split}\nonumber
   J_0&=i \mu^{4-d} \int \frac{ d^dq}{(2\pi)^d}\frac{ 1}{(q \cdot v-\omega+i \epsilon) (q^2-m^2_{\pi}+i \epsilon)}\\[2ex]\nonumber
&= \frac{\omega}{8 \pi^2}[1+\mathrm{R}-\mathrm{ln}(\frac{m_{\pi}^2}{\mu^2})-\frac{2}{\omega}F(\omega,m_{\pi})],
\end{split}
\end{equation}
 and the expression for $J_{\pi}$ is
\begin{align}\nonumber
   &J_\pi=i \mu^{4-d} \int \frac{ d^dq}{(2\pi)^d} \frac{1}{(q^2-m^2_{\pi}+i \epsilon)}= \frac{m^2_\pi}{16\pi^2}[\mathrm{ln}(\frac{m_{\pi}^2}{\mu^2})-\mathrm{R}],
\end{align}
where $\mathrm{R}=\frac{2}{4-d}-\gamma_E+\mathrm{ln}(4\pi)+1$ contains a pole
at $d=4$.
In these expressions, $\mu$ is the renormalization scale. The function $F(\omega,m_{\pi})$ is given in Eq.~\eqref{F-function}.

To use dimensional regularization consistently,
one has to set $d=4$ after expanding $J_2$ and $J_3$ to first order
in $4-d$. If one sets $d=4$ before expanding in powers of $4-d$ as in Refs. \cite{cas97,Grins},
the expressions for $J_2$ and $J_3$ will be missing some finite pieces
where  $\frac{1}{d-1}\mathrm{R}=\frac{1}{3}\mathrm{R}+\frac{2}{9}\neq \frac{1}{3}\mathrm{R}$.
If there is only one integral,
 then the different constants can be absorbed by
different renormalization schemes; i.e., this corresponds to some modified  subtraction schemes.
For the case of 
 two integrals with different finite terms,
 there is no single
consistent renormalization scheme; i.e., the differences cannot be hidden  in
renormalization schemes. 

By expanding  Eqs.~\eqref{eq39} and \eqref{eq40} to  first order in $4-d$ and then taking $d=4$, we get
\begin{equation}
\begin{split}
J_2=&\frac{1}{16 \pi^2} [(\frac{2}{3}\omega^3-m_{\pi}^2\omega)\mathrm{ln}(\frac{m_{\pi}^2}{\mu^2})+\frac{4}{3}(\omega^2- m_{\pi}^2)F(\omega,m_{\pi}) \\[2ex]
&-\frac{2}{3}\omega^3(\mathrm{R}+\frac{5}{3})+\frac{1}{3}\omega\,m_{\pi}^2 (3\mathrm{R}+4)],
\end{split}
\end{equation}
and
\begin{equation}
\begin{split}
J_3=&\frac{1}{16 \pi^2}[(2 m_{\pi}^2\omega-\frac{8}{3}\omega^3)\mathrm{ln}(\frac{m_{\pi}^2}{\mu^2})-\frac{4}{3}(4 \omega^2- m_{\pi}^2)F(\omega,m_{\pi}) \\[2ex]
&+\frac{8}{3}\omega^3(\mathrm{R}+\frac{7}{6})-\frac{2}{3}\omega\,m_{\pi}^2(3\mathrm{R}+2)].
\end{split}
\end{equation}

Now, by substituting Eq.~\eqref{eq02012} into Eq.~\eqref{chlointps00}, one gets
\begin{equation}
\begin{split}\label{chlointpsP0}
 i\,\Sigma^{(a)}_{H_1}&=3\left(\frac{g}{2\,f}\right)^2(g_{\mu \nu}-v_{\mu}v_{\nu})(-i\,(g^{\mu \nu} J_2+v^{\mu}v^{\nu}J_3))\\[2ex]
 &=3i \left(\frac{g}{2\,f}\right)^2 (1-g_{\mu \nu}g^{\mu \nu})J_2.
\end{split}
\end{equation}
As we have chosen to define the heavy meson  fields in four dimensions,
the contraction of the metric tensors is $g_{\mu \nu}g^{\mu \nu}=4$.
This
is  quite different from regularizing gauge theories in which the components of
the gauge boson fields are continued in $d$ dimensions to maintain the gauge invariance.
In contrast, here it is important that regularization
keeps the integrals
of  Figs. \ref{fig1}$(a)$, \ref{fig1}$(c)$, and \ref{fig1}$(d)$
equal. 
Our purpose  is to preserve the heavy quark symmetry.
As will be shown below, our choice of defining the meson
 field as four dimensional maintains this.

Thus, Eq.~\eqref{chlointpsP0} becomes
\begin{equation}
\begin{split}\label{chlointps0}
 i\,\Sigma^{(a)}_{H_1}&=3i\left(\frac{g}{2\,f}\right)^2 (-3\,J_2)=3i\left(\frac{g}{2\,f}\right)^2 K_1(\omega_a, m_{\pi}),
\end{split}
\end{equation}
where in the last step we introduced the chiral function $K_1(\omega, m_{\pi})$.
This can be related to $J_2$ as 
\begin{equation}
\begin{split}\label{kk1}
K_1(\omega, m_{\pi})=-3 J_2=&-\frac{3}{d-1}[(m^2_{\pi}-\omega^2)J_0-\omega J_{\pi}]\\[2ex]
=&\frac{1}{16 \pi^2}[(-2\omega^3+3m_{\pi}^2\omega)\mathrm{ln}(\frac{m_{\pi}^2}{\mu^2})-4(\omega^2- m_{\pi}^2)F(\omega,m_{\pi})\\[2ex]
&+2\omega^3(\mathrm{R}+\frac{5}{3})-\omega\, m_{\pi}^2 (3\mathrm{R}+4)],
\end{split}
\end{equation}
where this 
represents the contribution
to self-energy of charm mesons
from one-loop diagrams with  interacting  particles
belonging to the same doublets. 

Now, we want to calculate the integral of the loop diagram in
Fig.~\ref{fig1}$(c)$, which contributes to the self-energy of the vector charm meson
\begin{equation}
\begin{split}\label{chlointps10c}
 i\,\Sigma^{(c)}_{H^*_1}&=3\left(\frac{g}{2\,f}\right)^2\left(-\mu^{4-d} \int \frac{ d^dq}{(2\pi)^d} \frac{\epsilon \cdot q \epsilon \cdot q}{(q \cdot v-\omega_c+i \epsilon) (q^2-m^2_{\pi}+i \epsilon)}\right)\\[2ex]
&=3\left(\frac{g}{2\,f}\right)^2\left(-\epsilon^*_{\mu}\epsilon_{\nu}\mu^{4-d} \int \frac{ d^dq}{(2\pi)^d} \frac{q^{\mu} q^{\nu} }{(q \cdot v-\omega_c+i \epsilon) (q^2-m^2_{\pi}+i \epsilon)}\right)\\[2ex]
&=3i\left(\frac{g}{2\,f}\right)^2\epsilon^*_{\mu}\epsilon_{\nu}\,(g^{\mu \nu} J_2+v^{\mu}v^{\nu}J_3),\\
\end{split}
\end{equation}
where the last line is obtained by using Eq.~\eqref{eq02012}. Since $v_\mu \epsilon^\mu=0$ and $\epsilon^*_\mu \epsilon^\mu=-1$, $\Sigma^{(c)}_{H^*_1}$ is
\begin{equation}
\begin{split}\label{chlointps10c1}
 i\,\Sigma^{(c)}_{H^*_1}&=-3i\left(\frac{g}{2\,f}\right)^2\,J_2=i\left(\frac{g}{2\,f}\right)^2\,K_1(\omega_c, m_{\pi}).
\end{split}
\end{equation}

The  integral of one-loop diagram in Fig.~\ref{fig1}$(d)$, which contributes to the self-energy of the vector charm meson,
is
\begin{equation}
\begin{split}\label{chlointps10d}
 i\,\Sigma^{(d)}_{H^*_1}&=3\left(\frac{g}{2\,f}\right)^2\left(-\mu^{4-d} \int \frac{ d^dq}{(2\pi)^d} \frac{\epsilon^{\mu^{\prime}\nu^{\prime}\rho^{\prime}\sigma^{\prime}} \epsilon^*_{\mu^{\prime}}v_{\nu^{\prime}} q_{\rho^{\prime}}(g_{\sigma \sigma^{\prime}}-v_{\sigma^{\prime}} v_{\sigma})\epsilon^{\mu\nu\rho\sigma} \epsilon_{\mu}v_{\nu} q_{\rho}}{(q \cdot v-\omega_d+i \epsilon) (q^2-m^2_{\pi}+i \epsilon)}\right)\\[2ex]
&=-3i\left(\frac{g}{2\,f}\right)^2\epsilon^{\mu^{\prime}\nu^{\prime}\rho^{\prime}\sigma^{\prime}} \epsilon_{\mu^{\prime}\nu^{\prime}\rho^{\prime}\sigma^{\prime}}\epsilon^*_{\mu^{\prime}}\epsilon^{\mu^{\prime}}v_{\nu^{\prime}}v^{\nu^{\prime}} J_2.\\
\end{split}
\end{equation}
As $v \cdot v=1$, $\epsilon \cdot v=0$, and $\epsilon \cdot \epsilon=-1$,
the contraction between indices of the totally antisymmetric tensors yields  $-2!$.
Thus, $\Sigma^{(d)}_{H^*_1}$ becomes
\begin{equation}
\begin{split}
 i\,\Sigma^{(d)}_{H^*_1}&=3i\left(\frac{g}{2\,f}\right)^2 (-2 J_2)=2i\left(\frac{g}{2\,f}\right)^2 K_1(\omega_d, m_{\pi}).
\end{split}
\end{equation}
Clearly, our choice of defining meson fields
in four dimensions,
 which gives $g_{\mu\nu}g^{\mu\nu}=4$ for the loop integral
of  Fig.~\ref{fig1}$(a)$, yields results equal
to the loop integrals of Figs.~\ref{fig1}$(c)$ and \ref{fig1}$(d)$.
The results of the diagrams in Figs.~\ref{fig2}$(a)$, \ref{fig2}$(c)$, and \ref{fig2}$(d)$
are similar to the ones of Figs.~\ref{fig1}$(a)$, \ref{fig1}$(c)$, and \ref{fig1}$(d)$, respectively.

Now, we evaluate the loop integrals for graphs
describing the interaction of  heavy mesons with opposite parity.
To this end, let us begin with the second one-loop contribution to self-energy of $H_1$
which is shown in Fig.~\ref{fig1}$(b)$
\begin{equation}
\begin{split}\label{chlointps00b}
 i\,\Sigma^{(b)}_{H_1}&=3\left(\frac{h}{2\,f}\right)^2\left(-\mu^{4-d} \int \frac{ d^dq}{(2\pi)^d} \frac{v \cdot q  v \cdot q}{(q \cdot v-\omega_b+i \epsilon) (q^2-m^2_{\pi}+i \epsilon)}\right)\\[2ex]
&=3\left(\frac{h}{2\,f}\right)^2\left(-v_{\mu}v_{\nu}\mu^{4-d} \int \frac{ d^dq}{(2\pi)^d} \frac{q^{\mu} q^{\nu} }{(q \cdot v-\omega_b+i \epsilon) (q^2-m^2_{\pi}+i \epsilon)}\right).
\end{split}
\end{equation}
Similarly, substituting Eq.~\eqref{eq02012} into Eq.~\eqref{chlointps00b} gives
\begin{equation}
\begin{split}\label{chlointps00b*}
 i\,\Sigma^{(b)}_{H_1}&=3i\left(\frac{h}{2\,f}\right)^2v_{\mu}v_{\nu}\,(g^{\mu \nu} J_2+v^{\mu}v^{\nu}J_3)\\[2ex]
 &=3i\left(\frac{h}{2\,f}\right)^2(J_2+J_3)=3i\left(\frac{h}{2\,f}\right)^2\,K_2(\omega_b, m_{\pi}),
\end{split}
\end{equation}
where
\begin{equation}
\begin{split}\label{K2}
K_2(\omega, m_{\pi})=J_2+J_3=\omega^2J_0+\omega J_{\pi}=\frac{1}{16 \pi^2} [&(-2\omega^3+ m_{\pi}^2\omega)\mathrm{ln}(\frac{m_{\pi}^2}{\mu^2})-4\omega^2 F(\omega,m_{\pi})\\[2ex]
&+2\omega^3(1+\mathrm{R})- \omega\, m_{\pi}^2 \mathrm{R}].
\end{split}
\end{equation}
For the one-loop diagram with (heavy) interacting  particles
belonging to different doublets,
the contribution to the self-energy
is given by  the chiral function $K_2(\omega, m_{\pi})$.

The integral of the one-loop diagram shown in Fig.~\ref{fig1}$(e)$, which  contributes to the self-energy of the vector meson, is
\begin{equation}
\begin{split}\label{chlointps00e}
 i\,\Sigma^{(e)}_{H^*_1}&=3\left(\frac{h}{2\,f}\right)^2\mu^{4-d} \int \frac{ d^dq}{(2\pi)^d} \frac{\epsilon^*_{\mu}\,v \cdot q \,(g^{\mu \nu}-v^{\mu}v^{\nu})\, v \cdot q\, \epsilon_{\nu}}{(q \cdot v-\omega_e+i \epsilon) (q^2-m^2_{\pi}+i \epsilon)}\\[2ex]
&=3\left(\frac{h}{2\,f}\right)^2\epsilon^*_{\mu}\epsilon_{\nu}(g^{\mu \nu}-v^{\mu}v^{\nu})v_{\alpha} v_{\beta} \mu^{4-d} \int \frac{ d^dq}{(2\pi)^d} \frac{q^{\alpha} q^{\beta} }{(q \cdot v-\omega_e+i \epsilon) (q^2-m^2_{\pi}+i \epsilon)}\\[2ex]
&=-3\left(\frac{h}{2\,f}\right)^2\,v_{\alpha} v_{\beta} \mu^{4-d} \int \frac{ d^dq}{(2\pi)^d} \frac{q^{\alpha} q^{\beta} }{(q \cdot v-\omega_e+i \epsilon) (q^2-m^2_{\pi}+i \epsilon)}.\\
\end{split}
\end{equation}
Similarly, substituting Eq.~\eqref{eq02012} into Eq.~\eqref{chlointps00e} gives
\begin{equation}
\begin{split}\label{chlointps00b*}
 i\,\Sigma^{(e)}_{H^*_1}&=3i\left(\frac{h}{2\,f}\right)^2\,v_{\alpha} v_{\beta} (g^{\alpha \beta}J_2+v^{\alpha}v^{\beta}J_3)=3i\left(\frac{h}{2\,f}\right)^2 (J_2+J_3)\\[2ex]
 &=3i\left(\frac{h}{2\,f}\right)^2\,K_2(\omega_e, m_{\pi}).
\end{split}
\end{equation}

The loop integrals of the diagrams in Figs.~\ref{fig2}$(b)$ and \ref{fig2}$(e)$
are similar to the result of Figs.~\ref{fig1}$(b)$ and \ref{fig1}$(e)$, respectively.

\end{document}